\documentclass[superscriptaddress,showpacs,showkeys,amsmath,pre,preprint]{revtex4}
\usepackage{amssymb}
\usepackage{graphicx}% Include figure files
\usepackage{dcolumn}% Align table columns on decimal point
\usepackage{subfigure}
\usepackage{ifpdf}
\usepackage{bm}
\usepackage{multirow}

\begin{document}
\title{Path integral approach to the pricing of timer options with the Duru-Kleinert time transformation}
\author{L. Z. J. Liang}
\affiliation{TQC, Universiteit Antwerpen, Universiteitsplein 1,
2610 Antwerpen, Belgium}
\author{D. Lemmens}
\affiliation{TQC, Universiteit Antwerpen, Universiteitsplein 1,
2610 Antwerpen, Belgium}
\author{J. Tempere}
\affiliation{TQC, Universiteit Antwerpen, Universiteitsplein 1,
2610 Antwerpen, Belgium} \affiliation{Lyman Laboratory of Physics,
Harvard University, Cambridge, MA 02138.}

\date{\today}
\begin{abstract}
In this paper, a time substitution as used by Duru and Kleinert in their treatment of the hydrogen atom with path integrals is performed to price timer options under stochastic volatility models. We present general pricing formulas for both the perpetual timer call options and the finite time-horizon timer call options. These general results allow us to find closed-form pricing formulas for both the perpetual and the finite time-horizon timer options under the 3/2 stochastic volatility model as well as under the Heston stochastic volatility model. For the treatment of timer option under the 3/2 model we will rely on the path integral for the Morse potential, with the Heston model we will rely on the Kratzer potential.
\end{abstract}



\maketitle

\section {Introduction}\label{Introduction}
Timer options, first introduced for sale by Societe Generale Corporate and Investment Banking (SG CIB) in 2007 \cite{SG, Fontenay}, are relatively new products in the equity volatility market. The basic principle of this option is similar to the European vanilla option, with the key distinction being the uncertain expiration date. Rather than a fixed maturity time that is set at inception for the vanilla option, the expiry date of the timer option is a stopping time equal to the time needed for the realized variance of the underlying asset to reach a pre-specified level.

Stopping times, sometimes formulated as first passage or hitting times, have applications in various research fields. Traditional applications of stopping times in physics are for example the situation where internal fluctuations induce the current of an electric circuit to attain a critical value \cite{Roberts, Lindenberg} and Kramer's problem \cite{Weiss3, Hanggi}. Recent applications of stopping times can be found in neuroscience, where a neuron emits a signal when its membrane voltage exceeds a certain threshold \cite{Ditlevsen,Bulsara,Bibbona}; in the research field of quantum hitting times of Markov chains and hitting times of quantum random walks \cite{Krovi,Varbanov}; and in econophysics \cite{Masoliver,Katz}. For an introduction to first passage problems and an overview of possible applications see \cite{Redner,Weiss}.

When the expiration date is only determined by a stopping time that can theoretically become infinite the option is called a perpetual timer option. According to Hawkins and Krol \cite{Hawkins}, it is usual practice to specify a maximum expiry for the timer option, at which point the option expires in the same manner as vanilla options, to prevent excessively long maturity times. These options are called finite time-horizon timer options.

Timer options were first proposed in literature by Neuberger \cite{Neuberger} as "mileage" options in 1990. In the middle 1990s, Bick emphasized the application of dynamic trading strategies with timer options to portfolio insurance as well as to hedging strategies \cite{Bick}. Recently, after timer options were traded in the market, the amount of research concerning the pricing of perpetual timer options has increased. Li studied the pricing and hedging under the Heston stochastic volatility model \cite{Li}. Bernard and Cui proposed a fast and accurate almost-exact simulation method in general stochastic volatility models \cite{Bernard}. Saunders developed an asymptotic approximation under fast mean-reverting stochastic volatility models \cite{Saunders}. We contribute to the existing literature by presenting analytical pricing results for both perpetual and finite time-horizon timer options for a general stochastic process. These general results are then applied to determine explicit closed-form formulas for the 3/2 and the Heston stochastic volatility model. Especially for timer options it is relevant to investigate different stochastic volatility models, since the price of these options is particularly sensitive to the behavior of the volatility.

We will derive these results in the path integral framework. The applications of path integral methodology drawn from quantum mechanics to finance can be found, among others, in \cite{Kleinert, Linetsky, Bennati, Damiaan, Lingzhi, Baaquie} and references therein. The virtue of this method lies in its capacity to provide an intuitive way to derive the transition probability density function (propagator) of the underlying stochastic processes. Since the pricing of financial options comes down to evaluating expectation values of stochastic processes, the path integral formalism is a suitable alternative to partial differential equations.

To derive our general results we will rely on the Duru-Kleinert space-time substitution method used by Duru and Kleinert to treat the hydrogen atom with path integrals \cite{Kleinert,Duru, Duru2}. This method has recently been used in finance by Decamps and De Schepper to derive asymptotic formulas for Black-Scholes implied volatilities \cite{Decamps}. The Duru-Kleinert space-time substitution approach serves here to translate the original stochastic processes to new ones behaving in a stochastic time horizon. Under this new time horizon, the random expiry time is expressed as a functional of the transformed stochastic volatility. Then a method related to variational perturbation theory \cite{Feynman, Kleinert} is applied to derive the joint propagator of the transformed stochastic volatility process and the stopping time process in the new time horizon.  Based on these transition probability density functions, we arrive at the pricing formulas for the perpetual timer option. In addition, we obtain pricing formulas for the finite time-horizon timer option by deriving the joint propagator of the log-return and the realized variance process.

For stochastic volatility processes, we start by emphasizing the 3/2 stochastic volatility model \cite{Ahn, Andreasen, Gabriel} not only in view of its analytical tractability but also because of the support from empirical evidence \cite{Jones, Bakshi, Carr2}. The results for this model are obtained by making a connection with the Morse potential. Next we treat the Heston stochastic volatility model \cite{Heston} by relating it to the Kratzer potential. This leads to closed-form pricing formulas for perpetual and finite time-horizon timer options for both models. The result for the perpetual timer option under the Heston model corresponds to the one found by Li \cite{Li}, confirming our approach.

This paper is organized as follows. In section \ref{General} we present general pricing formulas for perpetual and finite time-horizon timer call options under general stochastic volatility models. Section \ref{Propagators} is devoted to deriving closed-form formulas for the 3/2 and the Heston stochastic volatility model. In section \ref{Disc} the closed-form formulas are compared with Monte Carlo simulations and some properties of timer options are discussed. And finally a conclusion is given in section \ref{Conclusion}.

\section{General pricing formula of timer call options}\label{General}
\subsection{Model description}
For conciseness of representation, in this paper we only consider option pricing in a risk-neutral world. Moreover we assume that the initial time of the option is the current time $t = 0$ because the generalization to the case of a forward-start option is straightforward.

Let $\{ S(t) \}$ denote the underlying asset price process following a Black-Scholes type stochastic differential equation (SDE), with a variance $v(t)$, which is stochastic variable itself. Conventionally, the time evolution of $S(t)$ is represented in terms of the log-return $x(t) = \ln \frac{S(t)}{S_0}$, with $S_0=S(0)$. The realization of a stochastic process $Z$ at a special time $s$ will be denoted by $Z_s$, and we will use this notation throughout the paper. After the transformation to the log-return, the system is governed by the SDEs:
\begin{align}
    d x (t) &= \left( r - \frac{v}{2} \right) dt + \sqrt{v} \left( \sqrt{1 - \rho^2} dW_1 +  \rho d W_2 \right), \label{dx} \\
    d v (t) &= \alpha(v) dt + \beta(v) d W_2, \label{dv}
\end{align}
where $r$ is the constant risk-neutral interest rate, $W_1(t)$ and $W_2(t)$ are two independent Wiener processes, $\rho \in [-1, 1]$ is the correlation coefficient between $x(t)$ and its variance $v(t)$.

Now we introduce the notion of the realized variance, which is a principal ingredient of timer options. In practice the realized variance is given by $ \sum \limits_{n=1}^{N} \left( x_{t_n} - x_{t_{n-1}}\right)^2 $, where the set of evaluation times $t_n$ are for example daily closing times. In the literature (see \cite{Sepp, Li}), the realized variance of the underlying asset during a time period $[0, T]$, denoted by $I_T$, is usually approximated by:
\begin{equation}
    I_T \, = \int_0^T v(t) dt. \label{defrv}
\end{equation}
Also in this paper equation (\ref{defrv}) will be used as the definition of the realized variance.

\subsection{Pricing of perpetual timer options}
The price of a perpetual timer call option with strike price $K$ can be expressed as the expectation of the discounted payoff:
\begin{equation} \label{Cperp1}
    \mathcal{C}_{Perp} = \mathbb{E} \left[ e^{- r \mathcal{T}_\mathcal{B}} \, \textrm{max} \left( S_0 \, e^{x_{\mathcal{T}_\mathcal{B}}} - K, 0 \right) \right].
\end{equation}
This expression is similar to the one for the vanilla call option, except for the uncertain expiry time $\mathcal{T}_\mathcal{B}$, which is the stopping time defined as
\begin{equation}
    \mathcal{T}_\mathcal{B} = \inf \left\{u > 0; \, \int_0^u v(t) dt = \mathcal{B} \right\}.
\end{equation}
Here $\mathcal{B} = \sigma_0^2 T_0$ is the pre-specified variance budget with $T_0$ the expected investment horizon and $\sigma_0$ the forecasted volatility of the underlying asset during that period.

The dependence on the implicitly defined expiry time $\mathcal{T}_\mathcal{B}$ is inconvenient. We will now apply the  Duru-Kleinert method of quantum mechanics \cite{Kleinert} to construct variables in function of which $\mathcal{T}_\mathcal{B}$ is explicitly given. Define a time substitution $\tau(t)$ such that
\begin{equation}
    \tau(t) = \int_0^t v(s) ds,
\end{equation}
we will refer to $\tau$ as the pseudotime, following \cite{Kleinert}. The inverse function theorem gives us that
\begin{equation}
    \frac{d \tau^{-1}(t)}{dt} = \frac{1}{v(\tau^{-1}(t))},
\end{equation}
from which it follows that $\tau^{-1}(t)$ is given by:
\begin{equation}
    \tau^{-1}(t) = \int_0^t \frac{1}{v(\tau^{-1}(s))} ds.
\end{equation}
Denote $v(\tau^{-1}(t))$ by $V(t)$ and $x(\tau^{-1}(t))$ by $X(t)$, which follow new SDEs:
\begin{align}
  d V(t) =& \frac{\alpha \left( V \right)}{V} dt + \frac{\beta\left( V \right)}{\sqrt{V}} d W_2, \\
  d X(t) =& \left( \frac{r}{V} - \frac{1}{2} \right) dt + \left( \sqrt{1 - \rho^2} dW_1 +  \rho d W_2 \right).
\end{align}
Given the timer call variance budget
\begin{equation}
    \mathcal{B} = \int_0^{\mathcal{T}_\mathcal{B}} v(t) dt = \tau(\mathcal{T}_\mathcal{B}),
\end{equation}
we obtain the explicit expression for the stopping time as
\begin{equation}\label{T}
    \mathcal{T}_\mathcal{B} = \tau^{-1}(\mathcal{B}) = \int_0^{\mathcal{B}} \frac{1}{v(\tau^{-1}(t))} dt = \int_0^{\mathcal{B}} \frac{1}{V(t)} dt.
\end{equation}
Note that $(x(t),v(t)) = (X(\tau (t) ),V(\tau (t) ))$ , so as $(x(t),v(t))$ evolves in the period $[0, \mathcal{T}_\mathcal{B}]$, $(X(t),V(t))$ evolves in $[ \tau (0), \tau (\mathcal{T}_\mathcal{B}) ]$, that is $[0, \mathcal{B}]$. Therefore $[ 0, \mathcal{B}]$ is now a fixed horizon in pseudotime, and not only do the processes $X$, $V$ and $\mathcal{T}$ evolve during that period, but also expression (\ref{Cperp1}) can be written as
\begin{equation}\label{Cperp2}
    \mathcal{C}_{Perp} = \mathbb{E} \left[ e^{- r \mathcal{T}_\mathcal{B}} \, \textrm{max} \left( S_0 \, e^{X_{\mathcal{B}}} - K, 0 \right) \right].
\end{equation}
Hence, it is intuitive to study the joint transition probability density function of the dynamics of $\left(X, \mathcal{T} \right)$. However, as $\mathcal{T}$ depends on $V$, we turn to the joint propagator of the dynamics of $\left(X, V, \mathcal{T} \right)$.

The substitutions
\begin{align}
  z(t) =& \int \frac{\sqrt{V}}{\beta \left( V \right)} d V(t), \label{z} \\
  y(t) =& X(t) - \rho z(t),
\end{align}
help change the correlated dynamics of $(X, V)$ into two independent processes following
\begin{align}
  dy(t) =& \left[ \frac{r}{V(z)} - \frac{1}{2} - \rho  \mathcal{A} \left( z \right) \right] dt + \sqrt{1 - \rho^2} d W_1, \\
  dz(t) =& \mathcal{A} \left( z \right) dt + d W_2,
\end{align}
where
\begin{equation} \label{A}
    \mathcal{A} \left( z (t) \right) = \frac{\alpha (V)}{\beta (V)  \sqrt{V}} + \frac{1}{2} \frac{d}{d V} \left( \frac{\sqrt{V}}{\beta (V) } \right) \frac{\beta^2 (V) }{V}
\end{equation}
is a function of $z(t)$ because $V(t)$ is expressed in terms of $z(t)$ according to expression (\ref{z}).

To determine the price of the timer option, the propagator $\mathcal{P} \left( y_\mathcal{B}, z_\mathcal{B}, \mathcal{T}_\mathcal{B} \, | \, y_0, z_0, 0  \right)$ is needed. This propagator describes the joint probability that $y$ has the value $y_\mathcal{B}$, $z$ has the value $z_\mathcal{B}$ and the stopping time has the value $\mathcal{T}_\mathcal{B}$ at a later pseudotime $\mathcal{B}$ given their initial value $y_0$, $z_0$ and 0 at pseudotime 0. Since the processes $y$ and $z$ are uncorrelated, the Lagrangian corresponding to their joint evolution can be written as $\mathcal{L}[y, \dot{y}, z]+\mathcal{L}[z, \dot{z}]$
with:
\begin{align}
  \mathcal{L}[y, \dot{y}, z] =& \frac{\left[ \dot{y} -  \left( \frac{r}{V(z)} - \frac{1}{2} - \rho  \mathcal{A} \left( z  \right) \right) \right]^2}{2 \left( 1 - \rho^2 \right)}, \\
  \mathcal{L}[z, \dot{z}] =& \frac{1}{2} \left[ \dot{z} - \mathcal{A} \left( z  \right) \right]^2 + \frac{1}{2} \frac{\partial}{\partial z} \mathcal{A} \left( z  \right),
\end{align}
Using the path integral framework, the joint propagator $\mathcal{P} \left( y_\mathcal{B}, z_\mathcal{B}, \mathcal{T}_\mathcal{B} \, | \, y_0, z_0, 0  \right)$ can be determined by:
\begin{align} \label{PyBzBTb}
  & \mathcal{P} \left( y_\mathcal{B}, z_\mathcal{B}, \mathcal{T}_\mathcal{B} \, | \, y_0, z_0, 0  \right) = \nonumber \\
  & \int \mathcal{D} y\int \mathcal{D} z  \delta \left( \mathcal{T}_\mathcal{B}  - \int_0^{\mathcal{B}} \frac{1}{V(z)} dt \right) \, e^{- \int_0^{\mathcal{B}} \left( \mathcal{L}[z, \dot{z}]+\mathcal{L}[y, \dot{y}, z] \right) dt},
\end{align}
where $\delta(\cdot)$ is the delta function. It serves here to select these paths of $V$, expressed in terms of $z$, such that $\int_0^{\mathcal{B}} \frac{1}{V(t)} dt$ equals $\mathcal{T}_\mathcal{B}$.

To proceed we introduce the Fourier transform of the delta function. Furthermore since the path integral corresponding to the $y$ variable is quadratic it can be solved analytically. After performing this path integral, we can return to the original $X_{\mathcal{B}}$ variable. Expression (\ref{PyBzBTb}) then becomes:
\begin{align}\label{PXzT}
    &  \mathcal{P} \left( X_{\mathcal{B}}, z_\mathcal{B}, \mathcal{T}_\mathcal{B} \, | \, y_0, z_0, 0  \right) = \nonumber \\
    & \int_{-\infty}^\infty \frac{d p}{2 \pi}  e^{i p  \mathcal{T}_\mathcal{B}} \int \mathcal{D} z \, e^{- \int_0^{\mathcal{B}} \left( \mathcal{L}[z, \dot{z}] + i p \frac{1}{V(z)} \right) dt} \frac{e^{- \frac{\left[ X_{\mathcal{B}} - \Upsilon \left( z \right)  \right]^2}{2  \left( 1 - \rho^2 \right) \mathcal{B}} }}{\sqrt{2 \pi \left( 1 - \rho^2 \right) \mathcal{B}}},
\end{align}
where
\begin{equation} \label{Upsilon}
    \Upsilon \left( z \right) = \rho \left( z_{\mathcal{B}} - z_0 - \int_0^{\mathcal{B}} \mathcal{A} \left( z (t) \right) dt \right) + r \mathcal{T}_\mathcal{B} - \frac{\mathcal{B}}{2}.
\end{equation}
In order to add the $z$ dependent term $\Upsilon \left(z\right)$ to the Lagrangian of the $z$ path integral one can introduce another Fourier integral, and $\mathcal{P} \left( X_{\mathcal{B}}, z_\mathcal{B}, \mathcal{T}_\mathcal{B} \, | \, y_0, z_0, 0  \right)$ then becomes:
\begin{align} \label{P3join}
  & \mathcal{P} \left( X_{\mathcal{B}}, z_\mathcal{B}, \mathcal{T}_\mathcal{B} \, | \, y_0, z_0, 0  \right) \nonumber \\
  = & \int_{-\infty}^\infty \frac{d l}{2\pi} \, e^{i \, l X_\mathcal{B} - \frac{\left( 1 - \rho^2 \right) \mathcal{B}}{2} l^2} \int_{- \infty}^\infty \frac{d p}{2 \pi} \, e^{i p \mathcal{T}_\mathcal{B} } \nonumber \\
  & \times \int \mathcal{D} z(t) \, e^{- \int_0^{\mathcal{B}} \left[ \mathcal{L}[z, \dot{z}] + i p \frac{1}{V(z)} + i l \Upsilon \left( z \right) \right] dt}.
\end{align}
Whether our approach will lead to closed-form pricing formulas for timer options will depend on the Lagrangian $ \mathcal{L}[z, \dot{z}] + i p \frac{1}{V(z)} + i l \Upsilon \left( z \right)$. More precisely this means that $\mathcal{A} \left( z (t) \right)$ and $\Upsilon \left( z \right)$ should be well behaved enough in terms of $z$. For the two examples illustrated in this paper, the functions $\alpha (V(t))$ and $\beta (V(t))$ are as such that $\Upsilon \left( z \right)$ is only a function of $z_\mathcal{B}$ and $\mathcal{T}_\mathcal{B}$, denoted by $\Upsilon \left(z_\mathcal{B}, \mathcal{T}_\mathcal{B} \right)$. Then it is not necessary to introduce the Fourier transform of expression (\ref{P3join}) and we can proceed with expression (\ref{PXzT}).
Now the price of a perpetual timer option which is given by
\begin{align} \label{CPerp0}
  \mathcal{C}_{Perp} =& \int dX_\mathcal{B} \int d z_\mathcal{B} \int d \mathcal{T}_\mathcal{B}  \,  \mathcal{P} \left( X_\mathcal{B}, z_\mathcal{B}, \mathcal{T}_\mathcal{B} \, | \, y_0, z_0, 0  \right) \nonumber \\
  & \times \left[ e^{- r \mathcal{T}_\mathcal{B}} \, \textrm{max} \left( S_0 \, e^{X_{\mathcal{B}}} - K, 0 \right) \right],
\end{align}
can be written as:
\begin{equation} \label{CPerp}
  \mathcal{C}_{Perp} = \int_0^\infty d \mathcal{T}_\mathcal{B} \int_{-\infty}^\infty d  z_{\mathcal{B}} \mathcal{P}\left( z_{\mathcal{B}}, \mathcal{T}_\mathcal{B} \, | \, z_0, 0  \right) \, \bar{\mathcal{C}} \left( z_{\mathcal{B}}, \mathcal{T}_\mathcal{B} \right),
\end{equation}
with $\bar{\mathcal{C}} \left( z_{\mathcal{B}}, \mathcal{T}_\mathcal{B} \right)$ being the prices conditional on $z$:
\begin{align}\label{C2}
  & \bar{\mathcal{C}} \left( z_{\mathcal{B}}, \mathcal{T}_\mathcal{B} \right) \nonumber \\
  =& \int_{-\infty}^\infty d X_{\mathcal{B}} \, \frac{e^{- \frac{\left[ X_{\mathcal{B}} - \Upsilon \left( z_{\mathcal{B}}, \mathcal{T}_\mathcal{B} \right)  \right]^2}{2  \left( 1 - \rho^2 \right) \mathcal{B}} }}{\sqrt{2 \pi \left( 1 - \rho^2 \right) \mathcal{B}}} \left[ e^{- r \mathcal{T}_\mathcal{B}} \, \textrm{max} \left( S_0 \, e^{X_{\mathcal{B}}} - K, 0 \right) \right] \nonumber \\
  =& S_0 \, e^{\Upsilon \left( z_{\mathcal{B}}, \mathcal{T}_\mathcal{B} \right) - r \mathcal{T}_\mathcal{B} + \frac{(1 - \rho^2) \mathcal{B}}{2} } \, \mathcal{N} \left( d_+ \right)  - K \, e^{- r \mathcal{T}_\mathcal{B} } \mathcal{N} (d_-),
\end{align}
where $\mathcal{N} (\cdot)$ is the cumulative distribution for the normal random variable and
\begin{align}
%  \bar{\Upsilon} \left( z_{\mathcal{B}}, \mathcal{T}_\mathcal{B} \right) &=& \rho \left( z_{\mathcal{B}} - z_0 - \int_0^{\mathcal{B}} \mathcal{A} \left( z (t) \right) dt \right) - \frac{\rho^2}{2} \mathcal{B}, \\
  d_+ =& \frac{\ln \frac{S_0}{K} + \left( 1 - \rho^2 \right) \mathcal{B} + \Upsilon \left( z_{\mathcal{B}}, \mathcal{T}_\mathcal{B} \right) }{\sqrt{\left( 1 - \rho^2 \right) \mathcal{B}}}, \\
  d_- =& \frac{\ln \frac{S_0}{K}  + \Upsilon \left( z_{\mathcal{B}}, \mathcal{T}_\mathcal{B} \right)}{\sqrt{\left( 1 - \rho^2 \right) \mathcal{B}}},
\end{align}
and $\mathcal{P}\left( z_{\mathcal{B}}, \mathcal{T}_\mathcal{B} \, | \, z_0, 0  \right)$ is given by
\begin{equation}\label{PXzT2}
     \mathcal{P} \left(z_\mathcal{B}, \mathcal{T}_\mathcal{B}  |  z_0, 0  \right) =
     \int_{-\infty}^\infty \frac{d p}{2 \pi}  e^{i p  \mathcal{T}_\mathcal{B}} \int \mathcal{D} z  e^{- \int_0^{\mathcal{B}} \left( \mathcal{L}[z, \dot{z}] + \frac{i p }{V(z)} \right) dt}.
\end{equation}
Note that expression (\ref{C2}) is a Black-Scholes-Merton type pricing formula for perpetual timer options. To determine the price of a perpetual timer option for a particular model one needs to evaluate $\Upsilon \left(z_\mathcal{B}, \mathcal{T}_\mathcal{B} \right)$ in order to obtain $\bar{\mathcal{C}} \left( z_{\mathcal{B}}, \mathcal{T}_\mathcal{B} \right)$. Furthermore if we also have the analytical expression for the joint propagator $\mathcal{P}\left( z_{\mathcal{B}}, \mathcal{T}_\mathcal{B} \, | \, z_0, 0  \right)$, formula (\ref{CPerp}) demonstrates that the closed-form perpetual timer option pricing formula can be derived through two trivial integrals.

\subsection{Pricing of finite time-horizon timer options}
In this subsection we consider the pricing of finite time-horizon timer option.

Let the maximum expiry time to be $T$, then the price of a finite time-horizon timer option, denoted by $\mathcal{C}_{Fini}$, of strike price $K$ can be expressed as a sum of two contributions:
\begin{equation}
    \mathcal{C}_{Fini} = \mathcal{C}_1 + \mathcal{C}_2,
\end{equation}
where
\begin{equation}
    \mathcal{C}_1 = \int_0^T d \mathcal{T}_\mathcal{B} \int_{-\infty}^\infty d  z_{\mathcal{B}} \mathcal{P}\left( z_{\mathcal{B}}, \mathcal{T}_\mathcal{B} \, | \, z_0, 0  \right) \, \bar{\mathcal{C}} \left( z_{\mathcal{B}}, \mathcal{T}_\mathcal{B} \right)
\end{equation}
is the contribution from paths that exhausted their variance budget before time $T$, and
\begin{equation}
    \mathcal{C}_2 = \, e^{- r T} \int_{-\infty}^\infty \left( S_0 \, e^{x_T} - K \right)_+ \mathcal{P}_\mathcal{B} \left( x_T \, | x_0 \, \right) dx_T
\end{equation}
is the contribution from paths that reach the preset finite time horizon. Note the integration range of $\mathcal{T}_\mathcal{B}$ in $\mathcal{C}_1$, which is truncated by the maximum expiry time $T$.

Denote the joint propagator of the log-return and the realized variance as $\mathcal{P} \left( x_T, I_T \, | \, x_0, 0  \right)$, then
\begin{equation} \label{PB}
    \mathcal{P}_\mathcal{B} \left( x_T \, | \, x_0  \right) = \int_0^\mathcal{B} \mathcal{P} \left( x_T, I_T \, | \, x_0, 0  \right) d I_T.
\end{equation}
Note $\mathcal{P}_\mathcal{B} \left( x_T \, | \, x_0  \right)$ is not the propagator of $x$ that should be used for the European vanilla option, which represents the probability that $x$ has the value $x_T$ at later time $T$ given the initial values $x_0$ at time $0$. Instead $\mathcal{P}_\mathcal{B} \left( x_T  \, | \, x_0  \right)$ in $\mathcal{C}_2$  is the propagator of $x$ which is also conditioned on the fact that the realized variance budget of each path has not been exhausted before the maximum expiry time $T$.

Furthermore, if $\mathcal{P}_\mathcal{B} \left( x_T \, | \, x_0  \right)$ can be written as a Fourier integral:
\begin{equation}
    \mathcal{P}_\mathcal{B} \left( x_T \, | \, x_0  \right) = \int_{- \infty}^\infty \frac{d l}{2 \pi} \, e^{i l \left( x_T - r T \right)} \mathcal{F} (l),
\end{equation}
then by following the derivation outlined in \cite{Kleinert2004}, we can rewrite $\mathcal{C}_2$ explicitly, and thus the pricing formula of finite time-horizon Timer option as
\begin{align} \label{CFini}
  \mathcal{C}_{Fini} =& \int_0^T d \mathcal{T}_\mathcal{B} \int_{-\infty}^\infty d  z_{\mathcal{B}} \mathcal{P}\left( z_{\mathcal{B}}, \mathcal{T}_\mathcal{B} \, | \, z_0, 0  \right) \, \bar{\mathcal{C}} \left( z_{\mathcal{B}}, \mathcal{T}_\mathcal{B} \right) \nonumber \\
  & +  \frac{\mathcal{G}(0)}{2} + i \int_{- \infty}^\infty \frac{d l}{2 \pi} \frac{e^{i l \left( \ln \frac{K}{S_0} - r T \right)} \mathcal{G}(l)}{l},
\end{align}
where
\begin{equation}
     \mathcal{G}(l) = S_0 \mathcal{F} (l + i) - K \, e^{- r T} \mathcal{F} (l).
\end{equation}

The integration of $\mathcal{P}_\mathcal{B} \left( x_T \,  | \, x_0  \right)$ over all possible $x_T$'s gives $\mathcal{F}(0)$, which is the "survival probability" describing the probability that the underling asset is executed at the maximum expiry time $T$. This survival probability can also be determined by $\int_T^\infty d \mathcal{T}_\mathcal{B} \int d z_\mathcal{B} \mathcal{P} \left( z_\mathcal{B}, \mathcal{T}_\mathcal{B} \, | \, z_0, 0 \right)$, from which it is clear that this probability is independent from the evolution of $x$, thus does not depend on the correlation coefficient $\rho$.

For finite time-horizon timer option, besides the evaluation of propagator $\mathcal{P}\left( z_{\mathcal{B}}, \mathcal{T}_\mathcal{B} \, | \, z_0, 0  \right)$ as for the perpetual timer option, we must also calculate the propagator $\mathcal{P} \left( x_T, I_T \, | \, x_0, 0  \right)$ to derive the formula of $\mathcal{F} (l)$.

\section{Propagators for the 3/2 and the Heston model}\label{Propagators}
In this section we focus on the derivations of joint propagators $\mathcal{P}\left( z_{\mathcal{B}}, \mathcal{T}_\mathcal{B} \, | \, z_0, 0  \right)$ and $\mathcal{P} \left( x_T, I_T \, | \, x_0, 0  \right)$. These are used in section \ref{Disc} in conjunction with expressions (\ref{CPerp}) and (\ref{CFini}) from the previous section to price perpetual and finite time-horizon timer options, respectively. Note that $\mathcal{P}\left( z_{\mathcal{B}}, \mathcal{T}_\mathcal{B} \, | \, z_0, 0  \right)$ is evaluated in the pseudotime horizon and $\mathcal{P} \left( x_T, I_T \, | \, x_0, 0  \right)$ in the original time horizon. The 3/2 and the Heston model are chosen both from mathematical and  empirical considerations.

As mentioned in the previous section, it is convenient to choose models such that $\int_0^\mathcal{B} \mathcal{A} \left( z(t) \right) dt$ is a function of $z_\mathcal{B}$ and $\mathcal{T}_\mathcal{B}$. In addition, from the perspective of mathematics, the total Lagrangian in expression (\ref{PXzT2}):
\begin{equation}
    \mathcal{L}_{Tot}[z, \dot{z}] = \mathcal{L}[z, \dot{z}] +  \frac{i p}{V(z)}
\end{equation}
written in terms of $z$ and $\dot{z}$ should be well behaved enough to achieve a closed-form solution with the path integral.

Furthermore there is substantial empirical evidence supporting the stochastic differential equation underlying the 3/2 model. The Heston model, on the other hand, is important because it is a standard model for the financial industry.

\subsection{The 3/2 model and the Morse potential}

The model dynamics of the 3/2 stochastic volatility model \cite{Ahn} is given by:
\begin{equation} \label{dv32}
    d v(t) = \kappa v \left( \theta - v \right) dt + \epsilon v^{3/2} dW_2.
\end{equation}
Relating this model to the general stochastic volatility model used in (\ref{dv}), we have
\begin{align}
    \alpha \left(V\right) =& \kappa V \left( \theta - V \right), \\
    \beta \left(V\right) =&  \epsilon V^{3/2}.
\end{align}
For calculation convenience, we multiply $z(t)$ defined in expression (\ref{z}) by a factor $- \epsilon$  to obtain
\begin{equation}
    z(t) =  - \ln V(t).
\end{equation}
Thus, according to equations (\ref{T}), (\ref{A}) and (\ref{Upsilon}), we have
\begin{eqnarray}
  \mathcal{T}_\mathcal{B}  &=&  \int_0^{\mathcal{B}} e^{z(t)} dt, \\
  \int_0^{\mathcal{B}} \mathcal{A} \left( z (t) \right) dt &=& \frac{\kappa \theta}{\epsilon} \mathcal{T}_\mathcal{B}  - \left( \frac{\kappa}{\epsilon} + \frac{\epsilon}{2} \right) \mathcal{B} \\
  \Upsilon \left( z_{\mathcal{B}}, \mathcal{T}_\mathcal{B} \right) &=& - \frac{\rho}{\epsilon} \left( z_{\mathcal{B}} + \ln v_0 \right) + r \mathcal{T}_\mathcal{B} - \frac{\mathcal{B}}{2} \nonumber \\
  && - \rho \left( \frac{\kappa \theta}{\epsilon} \mathcal{T}_\mathcal{B}  - \left( \frac{\kappa}{\epsilon} + \frac{\epsilon}{2} \right) \mathcal{B}  \right) , \label{Upsilon32}
\end{eqnarray}
Therefore the total Lagrangian is
\begin{align}
%  && \mathcal{L}_{Tota} [z, \dot{z}] \nonumber \\
%  &=&  \frac{1}{2 \, \epsilon^2} \left[ \dot{z} + \kappa \theta \, e^{z} - \left( \kappa + \frac{\epsilon^2}{2} \right) \right]^2 - \frac{\kappa \theta}{2}  \, e^{z} + i p \, e^{z} \nonumber \\
  \mathcal{L}_{Tot} [z, \dot{z}] &= \frac{1}{2 \epsilon^2} \dot{z}^2 + \frac{\kappa^2 \theta^2}{2 \epsilon^2} \, e^{2 z} - \left( \frac{\kappa^2 \theta}{\epsilon^2} + \kappa \theta - i p \right) \, e^z \nonumber \\
  & + \frac{\kappa \theta}{\epsilon^2} \, e^z \dot{z} - \left( \frac{\kappa}{\epsilon^2} + \frac{1}{2} \right) \dot{z} + \frac{\left( \kappa + \epsilon^2/2 \right)^2}{2 \epsilon^2}.
\end{align}
The nontrivial terms of $\mathcal{L}_{Tota}[z, \dot{z}]$ reveal that $z(t)$ is subjected to a Morse potential. By making use of the known path integral for the Morse potential \cite{Grosche}, see Appendix \ref{Morse}, the joint propagator is expressed as
\begin{align}
  & \mathcal{P} \left(z_{\mathcal{B}}, \mathcal{T}_\mathcal{B} \, | \, z_0, 0  \right) \nonumber \\
  =& \int_{- \infty}^\infty \frac{d p}{2 \pi} \, e^{i p \mathcal{T}_\mathcal{B} }  \int \mathcal{D} z(t) \, e^{- \int_0^{\mathcal{B}}  \mathcal{L}_{Tot}[z, \dot{z}]  dt} \nonumber \\
  =& e^{- \frac{\kappa \theta}{\epsilon^2} \left( e^{z_{\mathcal{B}}} - e^{z_0} \right)} \, e^{ \left( \frac{\kappa}{\epsilon^2} + \frac{1}{2} \right) (z_{\mathcal{B}} - z_0)} \, e^{- \frac{\left( \kappa + \frac{\epsilon^2}{2} \right)^2 \mathcal{B}}{2 \epsilon^2} } \int_{-\infty}^\infty \frac{d p}{2 \pi}  \nonumber \\
  & \times \, e^{i p \mathcal{T}_\mathcal{B}} \int \mathcal{D} z(t) \, e^{- \int_0^{\mathcal{B}} \left[ \frac{\dot{z}^2}{2 \epsilon^2}  + \frac{\kappa^2 \theta^2}{2 \epsilon^2} \, e^{2 z} - \left( \frac{\kappa^2 \theta}{\epsilon^2} + \kappa \theta - i p \right) \, e^z \right] dt} \nonumber \\
  =& \frac{\kappa \theta}{\epsilon^2 \sinh \frac{\kappa \theta \mathcal{T}_\mathcal{B}}{2}} \, e^{- \frac{\kappa \theta}{\epsilon^2} \left( e^{z_{\mathcal{B}}} - e^{z_0} \right) + \left( \frac{\kappa}{\epsilon^2} + \frac{1}{2} \right) \left( z_{\mathcal{B}} - z_0 \right) }   \nonumber \\
  & \times \, e^{- \left(\frac{\kappa}{\epsilon^2} + \frac{1}{2} \right)^2  \frac{\epsilon^2}{2} \mathcal{B} + \left( \frac{\kappa}{\epsilon^2} + 1 \right) \kappa \theta \mathcal{T}_\mathcal{B} - \frac{\kappa \theta}{\epsilon^2} \left( e^{z_\mathcal{B}} + e^{z_0} \right) \coth \frac{\kappa \theta \mathcal{T}_\mathcal{B}}{2}}   \nonumber  \\
  & \times \int_0^{\infty} \frac{d \Phi_I}{\pi} \textrm{Re} \left[\, e^{\Phi   \mathcal{B} }  \, I_{2 \sqrt{\frac{2}{\epsilon^2} \Phi}} \left( \frac{\frac{2 \kappa \theta}{\epsilon^2} \, e^{\frac{z_\mathcal{B} + z_0}{2}}}{\sinh \frac{\kappa \theta \mathcal{T}_\mathcal{B}}{2}} \right)  \right], \label{PzBTB32}
\end{align}
with $I_. \left( \cdot \right)$ the modified Bessel function of the first kind, and under the condition that the real part of the integration variable satisfies
\begin{equation}
    \Phi_R  >  \frac{2}{\epsilon^2} \left( \frac{\kappa}{\epsilon^2} + \frac{1}{2} \right)^2,
\end{equation}
according to (\ref{Restriction2}).

Plugging expressions (\ref{Upsilon32}) and (\ref{PzBTB32}) into formula (\ref{CPerp}) yields the closed-form pricing formula for the perpetual timer call options under the 3/2 model.

The integral over all possible $z_{\mathcal{B}}$ can be done analytically, which leads to the marginal propagator for the stopping time $\mathcal{T}_\mathcal{B}$:
\begin{align}
  \mathcal{P} \left(\mathcal{T}_\mathcal{B} \, | \, 0 \right)   =& \frac{\kappa \theta}{\epsilon^2} \left( 1 + \coth \frac{\kappa \theta \mathcal{T}_\mathcal{B}}{2} \right) \, e^{- \left( \frac{\kappa}{\epsilon^2} + \frac{1}{2} \right)^2  \frac{\epsilon^2}{2} \mathcal{B} }  \nonumber \\
  &  \times  \int_0^\infty \frac{d \Phi_I}{\pi} \, e^{\Phi  \mathcal{B}} \left( \frac{\mathcal{N}}{v_0} \right)^{\mathcal{M}} \frac{\Gamma \left( 2 \sqrt{\frac{2}{\epsilon^2} \Phi} - \mathcal{M} \right)}{\Gamma \left( 2 \sqrt{\frac{2}{\epsilon^2} \Phi} + 1 \right)} \nonumber \\
  & \times \, _1 F_1 \left( \mathcal{M} + 1; \, 2 \sqrt{\frac{2}{\epsilon^2} \Phi} + 1; \, - \frac{\mathcal{N}}{v_0} \right),
\end{align}
where $\Gamma \left( \cdot \right)$ is the Euler gamma function, $\, _1F_1 \left( \cdot; \cdot; \cdot \right)$ is the confluent hypergeometric function, and
\begin{align}
  \mathcal{N} (\mathcal{T}_\mathcal{B}) &=  \frac{\kappa \theta}{\epsilon^2} \left( \coth \frac{\kappa \theta \mathcal{T}_\mathcal{B}}{2} - 1 \right), \\
  \mathcal{M} \left( \Phi \right) &= \sqrt{\frac{2}{\epsilon^2} \Phi} - \left( \frac{\kappa}{\epsilon^2} + \frac{1}{2} \right).
\end{align}

\begin{figure*}
\centering
\includegraphics[width=3in]{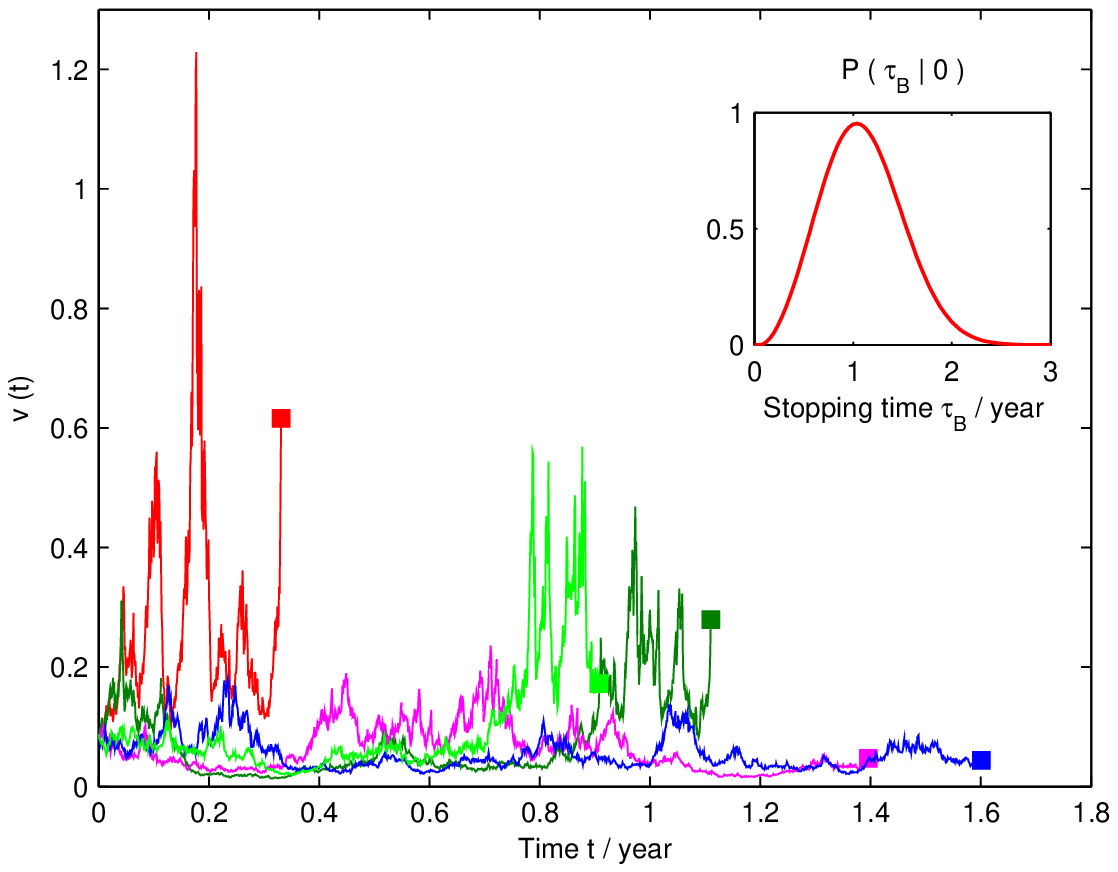}
\includegraphics[width= 3in]{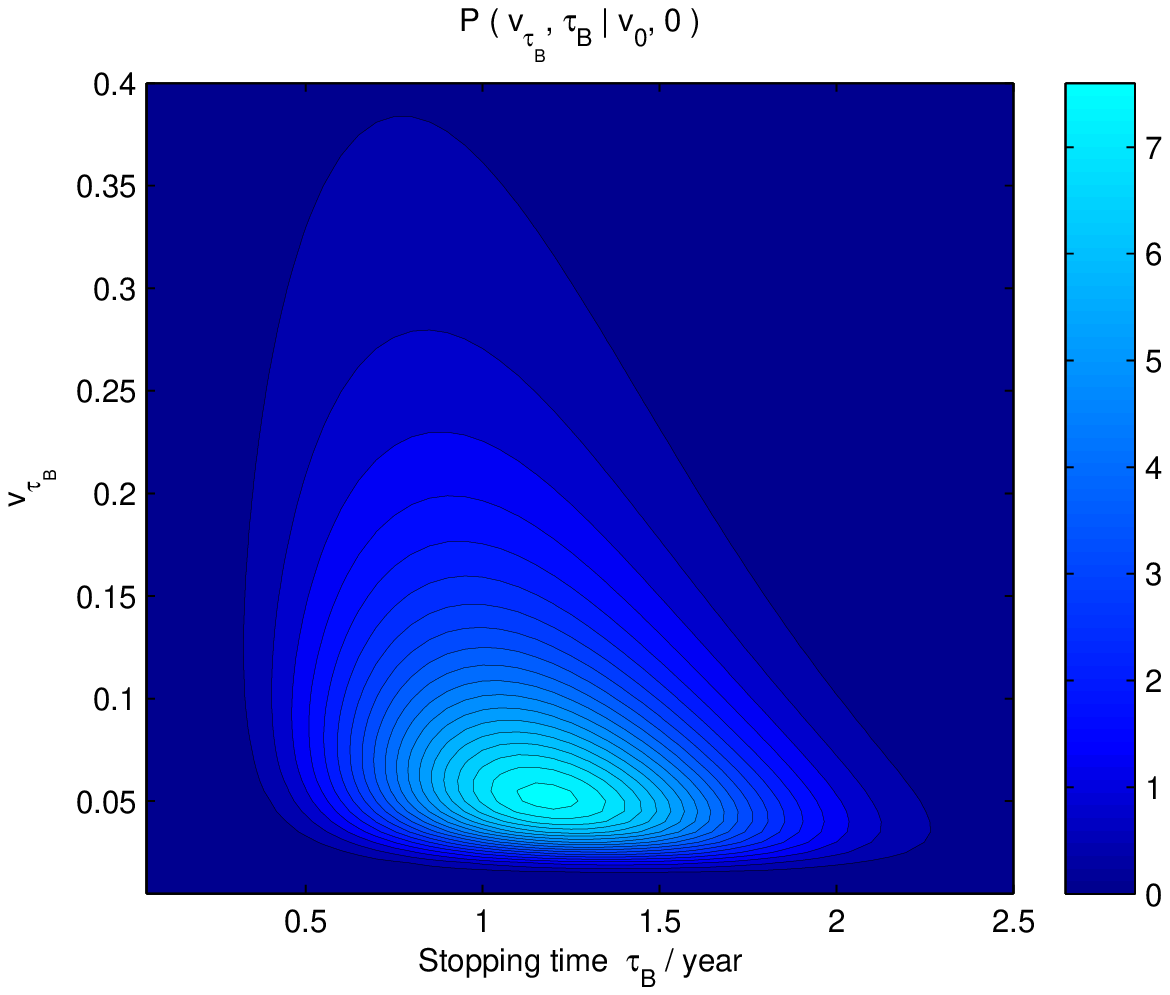}
\includegraphics[width=3in]{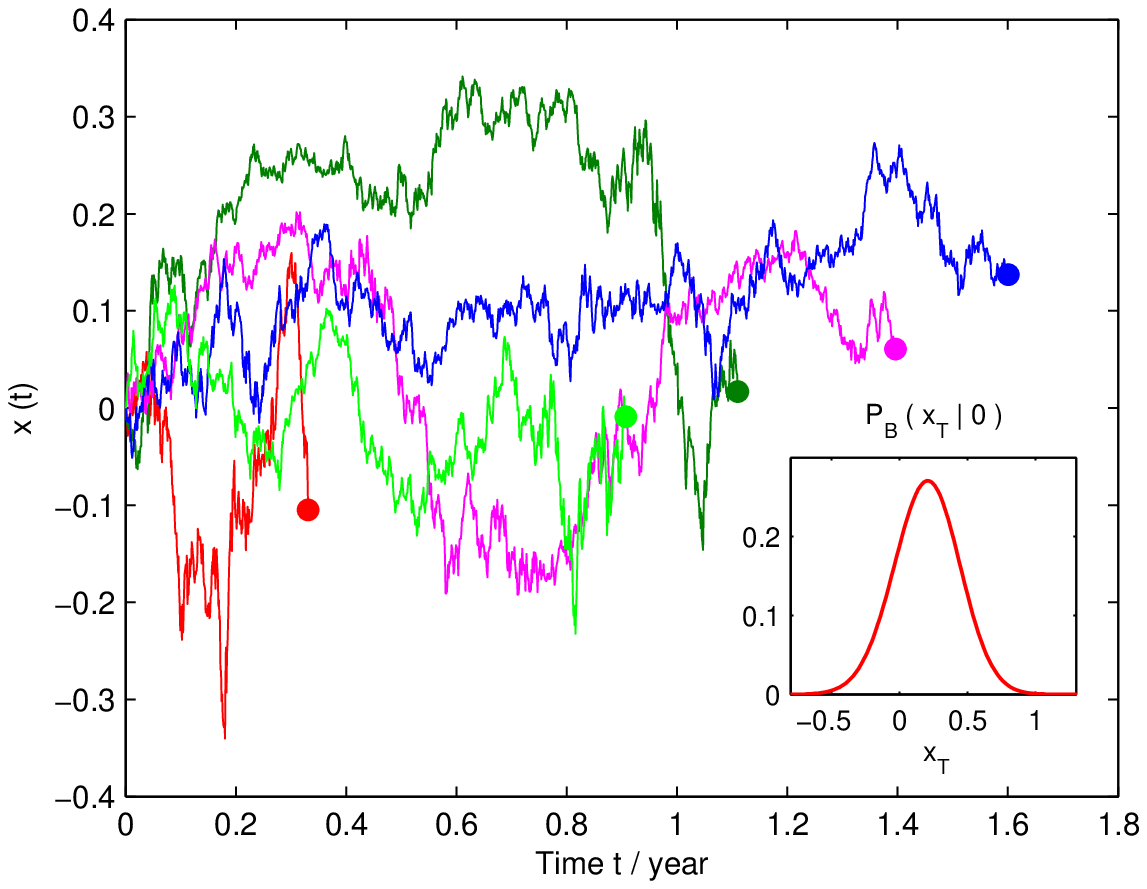}
\includegraphics[width= 3in]{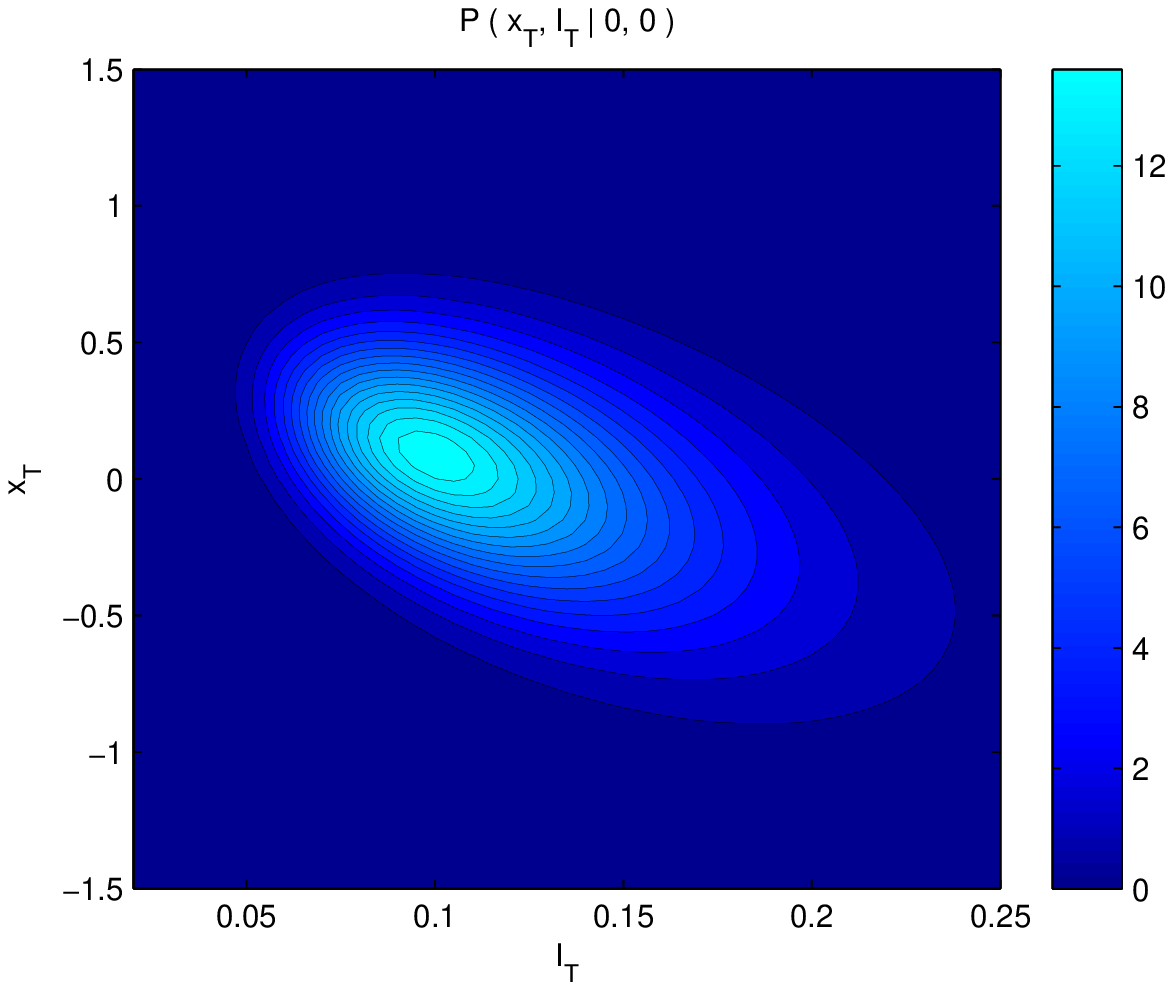} \\
\caption{(Color online, two columns) This figure shows several aspects of the time evolution of variables relevant for timer options under the 3/2 stochastic volatility model. The upper left panel shows several simulated variance paths up to the point where the realized variance reached $\mathcal{B}$. The bottom left panel shows corresponding log-return paths. The inset of the upper left panel shows the probability distribution of the stopping time $\mathcal{T}_\mathcal{B}$. The inset of the bottom left panel shows the density $\mathcal{P}_\mathcal{B} \left( x_T \, | \, x_0  \right)$ determined by expression (\ref{PB}). The upper right panel shows the joint probability distribution of the variance and the stopping time. The bottom right panel shows the joint probability distribution of the log-return and the realized variance. The parameters used here are: $v_0 = (0.295)^2, \kappa = 22.84, \theta = (0.4669)^2, \epsilon = 8.56, \mathcal{B} = v_0, r = 0.015, \rho = -0.5, T = 1.5$.} \label{fig1}
\end{figure*}

We move on to the calculation of the propagator $\mathcal{P} \left( x_T, I_T \, | \, x_0, 0  \right)$ by performing the following substitutions
\begin{align}
  \chi(t) & = x - \frac{\rho}{\epsilon} \left( \ln v - \kappa \theta t \right) - r t, \\
  \zeta(t) & = \frac{1}{\sqrt{v}},
\end{align}
which lead to two uncorrelated processes:
\begin{align}
  d \chi(t) =& \left( - \frac{1}{2} + \frac{\rho \kappa}{\epsilon} + \frac{\rho \, \epsilon}{2} \right) v dt + \sqrt{v} \sqrt{1 - \rho^2} dW_1 ,\\
  d \zeta(t) =&  \left[ - \frac{\kappa \theta}{2} \zeta + \left( \frac{\kappa}{2} + \frac{3}{8} \epsilon^2 \right) \frac{1}{\zeta} \right] dt - \frac{\epsilon}{2} dW_2,
\end{align}
and thus the corresponding Lagrangians:
\begin{align}
  \mathcal{L}[\chi, \dot{\chi}, v] &= \frac{\left[ \dot{\chi} - \left( - \frac{1}{2} + \frac{\rho \kappa}{\epsilon} + \frac{\rho \, \epsilon}{2} \right) v \right]^2}{2 v \left( 1 - \rho^2 \right)}  , \\
  \mathcal{L}[\zeta, \dot{\zeta}]
%  &= \frac{2}{\epsilon^2} \left[ \dot{\zeta} + \frac{\kappa \theta}{2} \zeta - \left( \frac{\kappa}{2} + \frac{3}{8} \epsilon^2 \right) \frac{1}{\zeta} \right]^2 \nonumber \\
%  & + \frac{1}{2} \left[ - \frac{\kappa \theta}{2} - \left( \frac{\kappa}{2} + \frac{3}{8} \epsilon^2 \right) \frac{1}{\zeta^2} \right]  \nonumber \\
  &= \mathcal{L}_1[\zeta] + \mathcal{L}_2[\zeta],
\end{align}
where
\begin{align}
  \mathcal{L}_1[\zeta, \dot{\zeta}] &= \frac{2}{\epsilon^2} \left[ \dot{\zeta}^2 + \frac{\kappa^2 \theta^2}{4} \zeta^2 \right] + \frac{\left( \frac{2 \kappa}{\epsilon^2} + 1 \right)^2 - \frac{1}{4}}{8 / \epsilon^2} \frac{1}{\zeta^2}, \\
  \mathcal{L}_2[\zeta, \dot{\zeta}] &= \frac{2 \kappa \theta}{\epsilon^2} \zeta \dot{\zeta} - \left( \frac{2 \kappa}{\epsilon^2} + \frac{3}{2} \right) \frac{\dot{\zeta}}{\zeta} - \left( \frac{\kappa^2 \theta}{\epsilon^2} + \kappa \theta \right).
\end{align}
Since $\chi$ is independent from $\zeta$, so the probability that $\chi$ goes to $\chi_T$, $\zeta$ goes to $\zeta_T$ and the realized variance reaches $I_T$ at a later time $T$ given the original positions $\chi_0$, $\zeta_0$ and $I_0=0$ at the initial time $0$ is

%Since $\chi$ is independent from $\zeta$, so the joint propagator of the dynamics of $\chi$, $\zeta$ and $I$ is
\begin{align}
  & \mathcal{P} (\chi_T, \zeta_T, I_T \, | \, \chi_0, \zeta_0, 0 ) \nonumber \\
  =& \int \mathcal{D} \zeta(t) \, \delta \left( I_T - \int_0^T v(t) \, dt \right) \, e^{- \int_0^T \mathcal{L}[\zeta, \dot{\zeta}] dt} \nonumber \\
  & \times \int \mathcal{D} \chi(t) \, e^{- \int_0^T \mathcal{L}[\chi, \dot{\chi}, v] dt} \nonumber \\
%  =& \int_{- \infty}^\infty \frac{d p}{2 \pi} \, e^{i p \,  I_T }  \int \mathcal{D} \zeta (t) \, e^{- i p \int_0^T v(t) \, dt } \, e^{- \int_0^T \mathcal{L}[\zeta, \dot{\zeta}] dt}   \nonumber \\
%  & \times \frac{e^{- \frac{\left[ x_T - \frac{\rho}{\epsilon}  \left( \ln v_T - \ln v_0 \right)  + \frac{\rho \kappa \theta}{\epsilon} T - r T +  \left( \frac{1}{2} - \frac{\rho \kappa}{\epsilon} - \frac{ \rho \, \epsilon}{2} \right) \int_0^T v(t) dt \right]^2}{2 \left( 1 - \rho^2 \right) \int_0^T v(t) dt}}}{\sqrt{2 \pi \left( 1 - \rho^2 \right) \int_0^T v(t) dt}}  \nonumber \\
  =& \left( \frac{\zeta_T}{\zeta_0} \right)^{\frac{2 \kappa}{\epsilon^2} + \frac{3}{2}} \, e^{- \frac{\kappa \theta}{\epsilon^2} (\zeta_T^2 - \zeta_0^2) + \left( \frac{\kappa^2 \theta}{\epsilon^2} + \kappa \theta \right) T}  \int_{- \infty}^\infty \frac{d p}{2 \pi} \, e^{i p \,  I_T } \nonumber \\
  & \times \int_{- \infty}^{+ \infty} \frac{d l}{2 \pi} \, e^{i l \left[ x_T  + \frac{\rho \kappa \theta}{\epsilon} T - r T \right]} \left( \frac{\zeta_T}{\zeta_0} \right)^{2 i l \frac{\rho}{\epsilon}} \nonumber \\
  & \times \int \mathcal{D} \zeta(t) \, e^{- \int_0^T \left[ \mathcal{L}_1[\zeta, \dot{\zeta}] +  \frac{ i l \left( - \frac{1}{2} + \frac{\rho \kappa}{\epsilon} + \frac{\rho \epsilon}{2} \right) + \frac{(1 - \rho^2) l^2}{2} + \, i p}{\zeta^2} \right] dt},
\end{align}
where the remaining path integral over $\mathcal{D} \zeta(t)$ of the radial harmonic oscillator potential \cite{Grosche} given by:
\begin{equation}
    \frac{2 \kappa \theta \sqrt{\zeta_T \zeta_0}}{\epsilon^2 \, \sinh (\frac{\kappa \theta T}{2}
    )} \, e^{- \frac{\kappa \theta}{\epsilon^2} \left(\zeta_T^2 + \zeta_0^2 \right) \coth (\frac{\kappa \theta T}{2})} \, I_\lambda \left( \frac{2 \kappa \theta \, \zeta_T \zeta_0}{\epsilon^2 \, \sinh (\frac{\kappa \theta T}{2})} \right),
\end{equation}
with
\begin{equation}
    \lambda = \left(
  \begin{array}{rl}
  & \left( \frac{2 \kappa}{\epsilon^2} + 1 \right)^2 \\
  & + \frac{8}{\epsilon^2} \left[ i l \left( - \frac{1}{2} + \frac{\rho \kappa}{\epsilon} + \frac{\rho \epsilon}{2} \right) + \frac{(1 - \rho^2) l^2}{2} + i p\right]
  \end{array}
  \right)^{\frac{1}{2}}.
\end{equation}
Integrating over $\zeta_T$ leads to $\mathcal{P} \left( \chi_T, I_T \, | \, \chi_0, 0  \right)$:
\begin{align} \label{chiTIT}
  & \mathcal{P} \left( \chi_T, I_T \, | \, \chi_0, 0  \right) \nonumber \\
  =& \int_0^\infty \mathcal{P} (\chi_T, \zeta_T, I_T \, | \, \chi_0, \zeta_0, 0 ) \, d \zeta_T  \nonumber \\
  =& \int_{- \infty}^\infty \frac{d p}{2 \pi} \, e^{i p \,  I_T } \int_{- \infty}^{+ \infty} \frac{d l}{2 \pi} \, e^{i l \left( x_T - r T \right)} \left( \frac{2}{\epsilon^2 N} \right)^M \nonumber \\
  & \times \frac{\Gamma(\lambda + 1 - M)}{\Gamma (\lambda + 1)} \, _1 F_1 \left( M; \lambda + 1; - \frac{2}{\epsilon^2 N} \right),
\end{align}
where
\begin{align}
  M & = \frac{\lambda}{2} - \frac{\kappa}{\epsilon^2} - \frac{1}{2} - i l \frac{\rho}{\epsilon}, \\
  N & = \frac{2 \sinh (\frac{\kappa \theta T}{2})}{\kappa \theta}  \, e^{\frac{\kappa \theta T}{2}} v_0.
\end{align}
Expression (\ref{chiTIT}) agrees with expression (73) in \cite{Carr2}.

According to (\ref{PB}), we have for the 3/2 model
\begin{equation} \label{PB32}
    \mathcal{P}_\mathcal{B} \left( x_T, T | x_0, 0 \right) = \int_{- \infty}^{+ \infty} \frac{d l}{2 \pi} \, e^{i l \left( x_T - r T \right)} \mathcal{F}(l),
\end{equation}
where
\begin{align} \label{Fl32}
  \mathcal{F}(l) = & - i \int_{- \infty}^\infty \frac{d p}{2 \pi} \frac{e^{i p \mathcal{B}} - 1}{p} \left( \frac{2}{\epsilon^2 N} \right)^M \frac{\Gamma(\lambda + 1 - M)}{\Gamma (\lambda + 1)} \nonumber \\
  & \times  \, _1 F_1 \left( M; \lambda + 1; - \frac{2}{\epsilon^2 N} \right).
\end{align}

The closed-form pricing formula for the finite time-horizon timer call options is derived by substituting (\ref{PB32}) and (\ref{Fl32}) in expression (\ref{CFini}).

\subsection{The Heston model and the Kratzer potential}
For Heston stochastic volatility model \cite{Heston}, the model dynamics is written as:
\begin{equation} \label{dvHes}
    d v(t) = \kappa \left( \theta - v \right) dt + \sigma \sqrt{v} dW_2.
\end{equation}
To relate this model to the general stochastic volatility model (\ref{dv}), $\alpha \left(V\right)$ and $\beta \left(V\right)$ are given by
\begin{align}
    \alpha (V) &= \kappa \left( \theta - V \right), \\
    \beta (V) & = \sigma \sqrt{V}.
\end{align}
From equation (\ref{z}), we have the relation between $z(t)$ and $V(t)$:
\begin{equation}
    z(t) = \frac{1}{\sigma} V(t).
\end{equation}
thus the stopping time $\mathcal{T}_\mathcal{B}$ is a functional of $z(t)$:
\begin{equation}
    \mathcal{T}_\mathcal{B} = \frac{1}{\sigma} \int_0^{\mathcal{B}} \frac{1}{z(t)} dt.
\end{equation}
Plugging equations (\ref{dvHes}) and (\ref{dv}) into definition (\ref{A}) gives
\begin{equation}
    \int_0^{\mathcal{B}} \mathcal{A} \left( z (t) \right) dt = \frac{\kappa \theta}{\sigma} \mathcal{T}_\mathcal{B} - \frac{\kappa}{\sigma} \mathcal{B},
\end{equation}
therefore (written in original variable $v_0$)
\begin{align}
  \Upsilon \left( z_{\mathcal{B}}, \mathcal{T}_\mathcal{B} \right) =& \rho \left( z_{\mathcal{B}} - \frac{v_0}{\sigma} - \frac{\kappa \theta}{\sigma} \mathcal{T}_\mathcal{B} + \frac{\kappa}{\sigma} \mathcal{B}  \right)+ r \mathcal{T}_\mathcal{B} - \frac{\mathcal{B}}{2}, \label{UpsilonHes} \\
  \mathcal{L}_{Tot}[z, \dot{z}]
%  =&  \frac{1}{2} \left[ \dot{z} - \left( \frac{\kappa \theta / \sigma^2}{z} - \frac{\kappa}{\sigma}   \right) \right]^2 - \frac{1}{2} \frac{\kappa \theta / \sigma^2}{z^2} +  \frac{i p}{\sigma} \frac{1}{z} \nonumber \\
  =& \frac{1}{2} \dot{z}^2 + \frac{\lambda^2 - \frac{1}{4}}{2 z^2} - \frac{\left( \lambda + \frac{1}{2} \right) \mu - \frac{i p}{\sigma}}{z} \nonumber \\
  & - \left( \lambda + \frac{1}{2} \right) \frac{\dot{z}}{z} + \mu \dot{z} + \frac{1}{2} \mu^2,
\end{align}
where
\begin{equation}
    \lambda = \frac{\kappa \theta}{\sigma^2} - \frac{1}{2}, \quad \quad \mu = \frac{\kappa}{\sigma}.
\end{equation}
The nontrivial terms of the total Lagrangian $\mathcal{L}_{Tot}[z, \dot{z}]$ manifest that $z(t)$ is subjected to a Kratzer potential. With the help of the known path integral for Kratzer potential \cite{Grosche}, see Appendix \ref{Kratzer}, we obtain the joint propagator as:
\begin{align}
  & \mathcal{P} \left(z_{\mathcal{B}}, \mathcal{T}_\mathcal{B} \, | \, z_0, 0 \right) \nonumber \\
  =& \int \mathcal{D} z(t) \, \delta \left( \mathcal{T}_\mathcal{B} - \frac{1}{\sigma} \int_0^{\mathcal{B}} \frac{1}{z(t)} dt \right) \, e^{- \int_0^{\mathcal{B}} \mathcal{L} [z, \dot{z}] dt} \nonumber \\
  =& \left( \frac{z_{\mathcal{B}}}{z_0} \right)^{\lambda + \frac{1}{2}} \, e^{- \mu \left( z_{\mathcal{B}} - z_0 \right) - \frac{1}{2} \mu^2 \mathcal{B} } \int_{-\infty}^\infty \frac{d p}{2 \pi} \, e^{i p \mathcal{T}_\mathcal{B}} \nonumber \\
  & \times \int \mathcal{D} z(t) \, e^{- \int_0^{\mathcal{B}} \left[ \frac{1}{2} \dot{z}^2 + \frac{\lambda^2 - \frac{1}{4}}{2 z^2} - \frac{\left( \lambda + \frac{1}{2}\right) \mu -  \frac{i p}{\sigma}}{z} \right] dt} \nonumber \\
  =& \frac{\sigma}{2} \, \frac{z_{\mathcal{B}}^{\lambda + 1}}{z_0^\lambda} \, e^{- \mu \left( z_{\mathcal{B}} - z_0 \right) - \frac{1}{2} \mu^2 \mathcal{B} + \left( \lambda + \frac{1}{2} \right) \mu \sigma \mathcal{T}_\mathcal{B}} \int_0^\infty \frac{d \Phi_I}{\pi}  \nonumber \\
  & \times  \textrm{Re} \left[
  \begin{array}{rl}
  &  \, e^{\Phi \mathcal{B}}  \, e^{ - \sqrt{2 \Phi} \left( z_{\mathcal{B}} + z_0 \right) \coth \left( \sqrt{\frac{\Phi}{2}} \sigma \mathcal{T}_\mathcal{B}\right)}  \\
  & \times \frac{2 \sqrt{2 \Phi}}{\sinh \left( \sqrt{\frac{\Phi}{2}} \sigma \mathcal{T}_\mathcal{B} \right)} \, I_{2 \lambda} \left( \frac{2 \sqrt{2 \Phi} \sqrt{z_{\mathcal{B}} z_0}}{\sinh \left( \sqrt{\frac{\Phi}{2}} \sigma \mathcal{T}_\mathcal{B} \right)} \right)
  \end{array}
  \right] \label{JoinH1}.
\end{align}
According to the condition given by (\ref{Restriction1}), the real part of the complex variable $\Phi$ must satisfy
\begin{equation}
    \Phi_R >  \frac{\kappa^2}{2 \, \sigma^2}.
\end{equation}
Again, plugging the expressions (\ref{UpsilonHes}) and (\ref{JoinH1}) into formula (\ref{CPerp}) yields the closed-form pricing formula for the perpetual timer call options under the Heston model .

Note the similarity of expression (\ref{JoinH1}) with the result obtained in \cite{Li}. Expression (\ref{JoinH1}) has a clear meaning as joint transition probability density function which illustrates the conciseness of physics; path integration allows to derive expression ($\ref{JoinH1}$) without applying any previous results of Bessel processes as done in \cite{Li}.

Li \cite{Li} computed the risk-neutral expected maturity in expression (5.2) by doing two numerical integrals. Actually, we can derive the marginal propagator of the stopping time $\mathcal{T}_\mathcal{B}$ by integrating over all possible $z_{\mathcal{B}}$ as follows
\begin{align}
  &  \mathcal{P} \left(\mathcal{T}_\mathcal{B} \, | \, 0 \right) \nonumber \\
  =& \int_0^\infty \mathcal{P} \left(z_{\mathcal{B}}, \mathcal{T}_\mathcal{B} \, | \, z_0, 0  \right) d z_{\mathcal{B}} \nonumber \\
  =& \sigma \, e^{- \frac{1}{2} \mu^2 \mathcal{B} + \left( \lambda + \frac{1}{2} \right) \mu \sigma \mathcal{T}_\mathcal{B}} \int_0^\infty \frac{d \Phi_I}{\pi} \nonumber \\
  & \times \textrm{Re} \left[
  \begin{array}{rl}
  & \,  \exp\left\{\Phi \mathcal{B} + \frac{ \left(\mu^2 - 2 \Phi \right) \, z_0}{\mu + \sqrt{2 \Phi} \coth \left( \sqrt{\frac{\Phi}{2}} \, \sigma \mathcal{T}_\mathcal{B} \right)}  \right\} \\
  & \times \,   \frac{\left( \frac{\sqrt{2 \Phi} }{\sinh \left( \sqrt{\frac{\Phi}{2}} \sigma \mathcal{T}_\mathcal{B} \right)} \right)^{2 \lambda +  1}}{\left( \mu + \sqrt{2 \Phi} \coth \left( \sqrt{\frac{\Phi}{2}} \, \sigma \mathcal{T}_\mathcal{B} \right) \right)^{2 \lambda + 2}}\\
  & \times \left( 2 \lambda + 1 + \frac{\left( \frac{\sqrt{2 \Phi}}{\sinh \left( \sqrt{\frac{\Phi}{2}} \sigma \mathcal{T}_\mathcal{B} \right)} \right)^{2} z_0}{\mu + \sqrt{2 \Phi} \coth \left( \sqrt{\frac{\Phi}{2}} \, \sigma \mathcal{T}_\mathcal{B} \right)} \right)
  \end{array}
  \right].
\end{align}

For the calculation of $\mathcal{P} \left( x_T, I_T |  0, 0  \right)$, we follow the derivation in \cite{Damiaan}. Substitutions
\begin{align}
  \chi(t) &= x - \frac{\rho}{\sigma} \left( v - \kappa \theta t \right) - r t, \\
  \zeta(t) &= \sqrt{v},
\end{align}
give two uncorrelated processes:
\begin{align}
  d \chi(t) &= \left( \frac{\rho \kappa}{\sigma}  - \frac{1}{2} \right) v  \, dt + \sqrt{v } \sqrt{1 - \rho^2} dW_1, \\
  d \zeta(t) &= \left[ \frac{\kappa \theta - \frac{\sigma^2}{4}}{2 \zeta} - \frac{\kappa}{2} \zeta \right] dt + \frac{\sigma}{2} dW_2.
\end{align}
The corresponding Lagrangians are:
\begin{align}
  \mathcal{L} \left[ \chi, \dot{\chi}, v \right] &= \frac{1}{2 v \left( 1 - \rho^2 \right)} \left[ \dot{\chi} - \left( \frac{\rho \kappa}{\sigma}  - \frac{1}{2} \right) v \right]^2, \\
  \mathcal{L} [ \zeta, \dot{\zeta} ]
%  &=& \frac{2 \left[ \dot{\zeta} - ( \frac{\kappa \theta - \frac{\sigma^2}{4}}{2 \zeta} - \frac{\kappa}{2} \zeta ) \right]^2}{\sigma^2}  - \frac{\kappa \theta - \frac{\sigma^2}{4}}{4 \zeta^2} - \frac{\kappa}{4} \nonumber \\
  &= \mathcal{L}_1 [ \zeta, \dot{\zeta} ] + \mathcal{L}_2 [ \zeta, \dot{\zeta} ],
\end{align}
where
\begin{align}
  \mathcal{L}_1 \left[ \zeta, \dot{\zeta} \right] =& \frac{2}{\sigma^2} \dot{\zeta}^2 + \frac{(\kappa \theta - \frac{\sigma^2}{4})(\kappa \theta - \frac{3 \sigma^2}{4})}{2 \, \sigma^2 \zeta^2}  + \frac{\kappa^2}{2 \sigma^2} \zeta^2, \\
  \mathcal{L}_2 \left[ \zeta, \dot{\zeta} \right] =&  - \left( \frac{2 \kappa \theta }{\sigma^2} - \frac{1}{2} \right) \frac{\dot{\zeta}}{\zeta} + \frac{2 \kappa}{\sigma^2} \zeta \dot{\zeta} - \frac{\kappa^2 \theta}{\sigma^2}.
\end{align}

Since $\chi$ is independent from $\zeta$, we similarly have the joint propagator of the dynamics of $\chi$, $\zeta$ and $I$:
\begin{align}
  & \mathcal{P} (\chi_T, \zeta_T, I_T \, | \, \chi_0, \zeta_0, 0 ) \nonumber \\
  =& \int \mathcal{D} \zeta(t) \, \delta \left( I_T - \int_0^T v(t) \, dt \right) \, e^{- \int_0^T \mathcal{L}[\zeta, \dot{\zeta}] dt} \nonumber \\
  & \times \int \mathcal{D} \chi(t) \, e^{- \int_0^T \mathcal{L}[\chi, \dot{\chi}, v] dt} \nonumber \\
%  =& \int_{- \infty}^\infty \frac{d p}{2 \pi} \, e^{i p \,  I_T }  \int \mathcal{D} \zeta(t) \, e^{- i p \int_0^T v(t) \, dt } \, e^{- \int_0^T \mathcal{L}[\zeta, \dot{\zeta}] dt}   \nonumber \\
%  & \times \frac{e^{- \frac{\left[ x_T - \frac{\rho}{\sigma}  \left( v_T - v_0 \right)  + \frac{\rho \kappa \theta}{\sigma} T - r T -  \left(  \frac{\rho \kappa}{\sigma} - \frac{1}{2} \right) \int_0^T v(t) dt \right]^2}{2 \left( 1 - \rho^2 \right) \int_0^T v(t) dt}}}{\sqrt{2 \pi \left( 1 - \rho^2 \right) \int_0^T v(t) dt}}  \nonumber \\
  =& \left( \frac{\zeta_T}{\zeta_0} \right)^{\frac{2 \kappa \theta}{\sigma^2} - \frac{1}{2}} \, e^{- \frac{\kappa}{\sigma^2} (\zeta_T^2 - \zeta_0^2) +  \frac{\kappa^2 \theta}{\sigma^2}  T}  \int_{- \infty}^\infty \frac{d p}{2 \pi} \, e^{i p \,  I_T } \nonumber \\
  & \times \int_{- \infty}^{+ \infty} \frac{d l}{2 \pi} \, e^{i l \left[ x_T  + \frac{\rho \kappa \theta}{\sigma} T - r T \right]} \, e^{- i l \frac{\rho}{\sigma} \left(\zeta_T^2 - \zeta_0^2 \right)} \nonumber \\
  & \times \int \mathcal{D} \zeta(t) \, e^{- \int_0^T \left[ \mathcal{L}_1[\zeta, \dot{\zeta}] +  \left( i l \left( \frac{\rho \kappa}{\sigma} - \frac{1}{2}  \right) +  \frac{\left(1 - \rho^2\right) l^2}{2}  + i p \right) \zeta^2 \right] dt},
\end{align}
where the path integral for the radial harmonic oscillator potential \cite{Grosche} is given by:
\begin{equation}
    \frac{4 \omega \sqrt{\zeta_T \zeta_0}}{\sigma^2 \, \sinh (\omega T)} \, e^{- \frac{2 \omega}{\sigma^2} \left(\zeta_T^2 + \zeta_0^2 \right) \coth (\omega T)} \, I_{\frac{2 \kappa \theta}{\sigma^2} - 1} \left( \frac{4 \omega \, \zeta_T \zeta_0}{\sigma^2 \, \sinh (\omega T)} \right),
\end{equation}
where
\begin{equation}
    \omega = \frac{\sigma}{2} \sqrt{ \frac{\kappa^2}{\sigma^2} + \left(1 - \rho^2\right) l^2 + i l \left( \frac{2 \rho \kappa}{\sigma} - 1\right) + 2 i p}.
\end{equation}
Integrating over $\zeta_T$ leads to $\mathcal{P} \left( \chi_T, I_T \, | \, \chi_0, 0  \right)$:
\begin{align} \label{chiTIT2}
  & \mathcal{P} \left( \chi_T, I_T \, | \, \chi_0, 0  \right) \nonumber \\
  =& \int_0^\infty \mathcal{P} (x_T, \zeta_T, I_T \, | \, x_0, \zeta_0, 0 ) \, d \zeta_T  \nonumber \\
  =& \, e^{\frac{\kappa}{\sigma^2} v_0 + \frac{\kappa^2 \theta}{\sigma^2} T} \int_{- \infty}^\infty \frac{d p}{2 \pi} \, e^{i p \,  I_T } \int_{- \infty}^{ \infty} \frac{d l}{2 \pi} \, e^{i l \left( x_T - r T \right)}  \nonumber \\
  & \times \, e^{i l \frac{\rho}{\sigma} \left( \kappa \theta T + v_0 \right)} N^{\frac{2 \kappa \theta}{\sigma^2}} \, e^{- \frac{2 \omega \left( \cosh (\omega T) - N \right)}{\sigma^2 \sinh (\omega T)} \, v_0},
\end{align}
where
\begin{equation}
    N = \left( \cosh (\omega T) + \frac{\kappa + i l \rho \sigma}{2 \omega} \sinh (\omega T) \right)^{-1}.
\end{equation}
Note the similarity of expression (\ref{chiTIT2}) with the result obtained in \cite{Gabriel}.

According to (\ref{PB}), we have for the Heston model
\begin{equation}
    \mathcal{P}_\mathcal{B} \left( x_T \, | \, x_0  \right) = \int_{- \infty}^{+ \infty} \frac{d l}{2 \pi} \, e^{i l \left( x_T - r T \right)} \mathcal{F}(l),
\end{equation}
where
\begin{eqnarray}
  \mathcal{F}(l) &=& - i \, e^{\frac{\kappa}{\sigma^2} v_0 + \frac{\kappa^2 \theta}{\sigma^2} T} \int_{- \infty}^\infty \frac{d p}{2 \pi} \frac{e^{i p \mathcal{B}} - 1}{p} \nonumber \\
  && \times \, e^{i l \frac{\rho}{\sigma} \left( \kappa \theta T + v_0 \right)} N^{\frac{2 \kappa \theta}{\sigma^2}} \, e^{- \frac{2 \omega \left( \cosh (\omega T) - N \right)}{\sigma^2 \sinh (\omega T)} \, v_0},
\end{eqnarray}
with which we obtain the closed-form pricing formula for the finite time-horizon timer call options according to formula (\ref{CFini}).

\begin{figure*}
\centering
\includegraphics[width=3in]{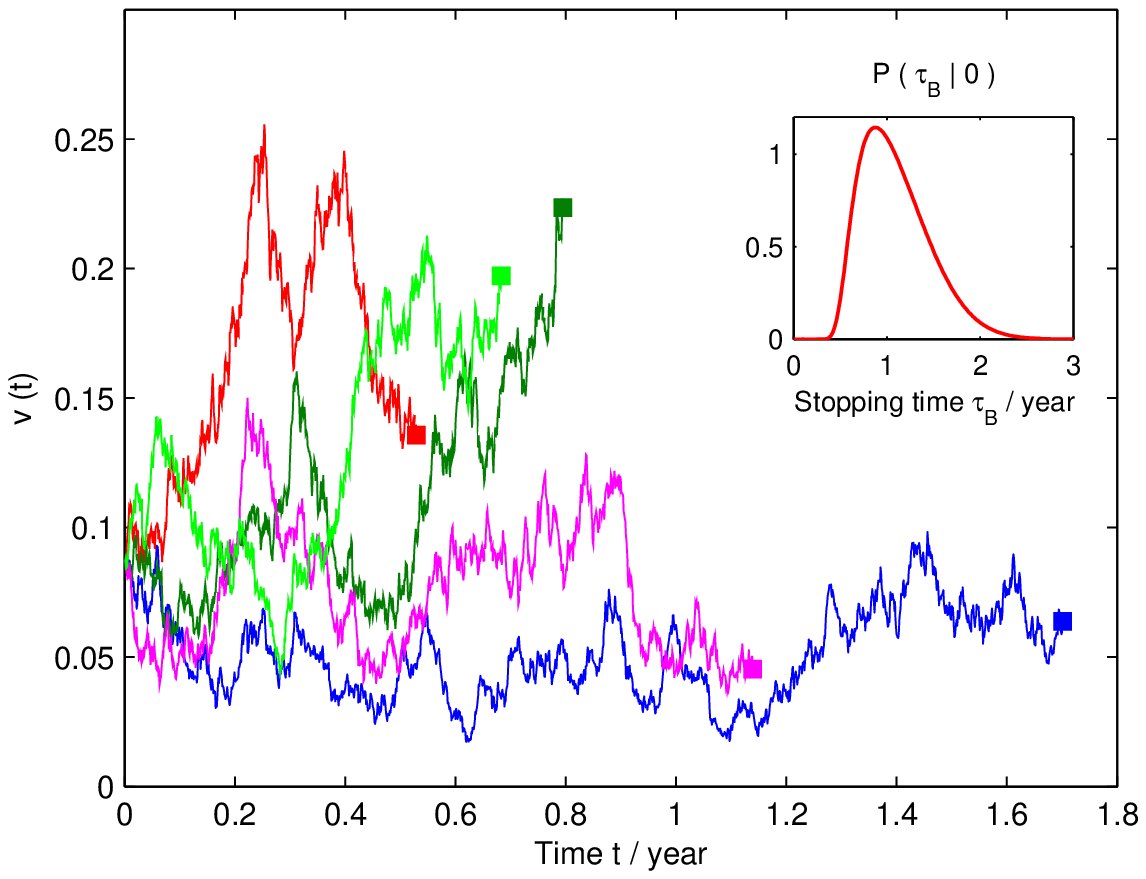}
\includegraphics[width= 3in]{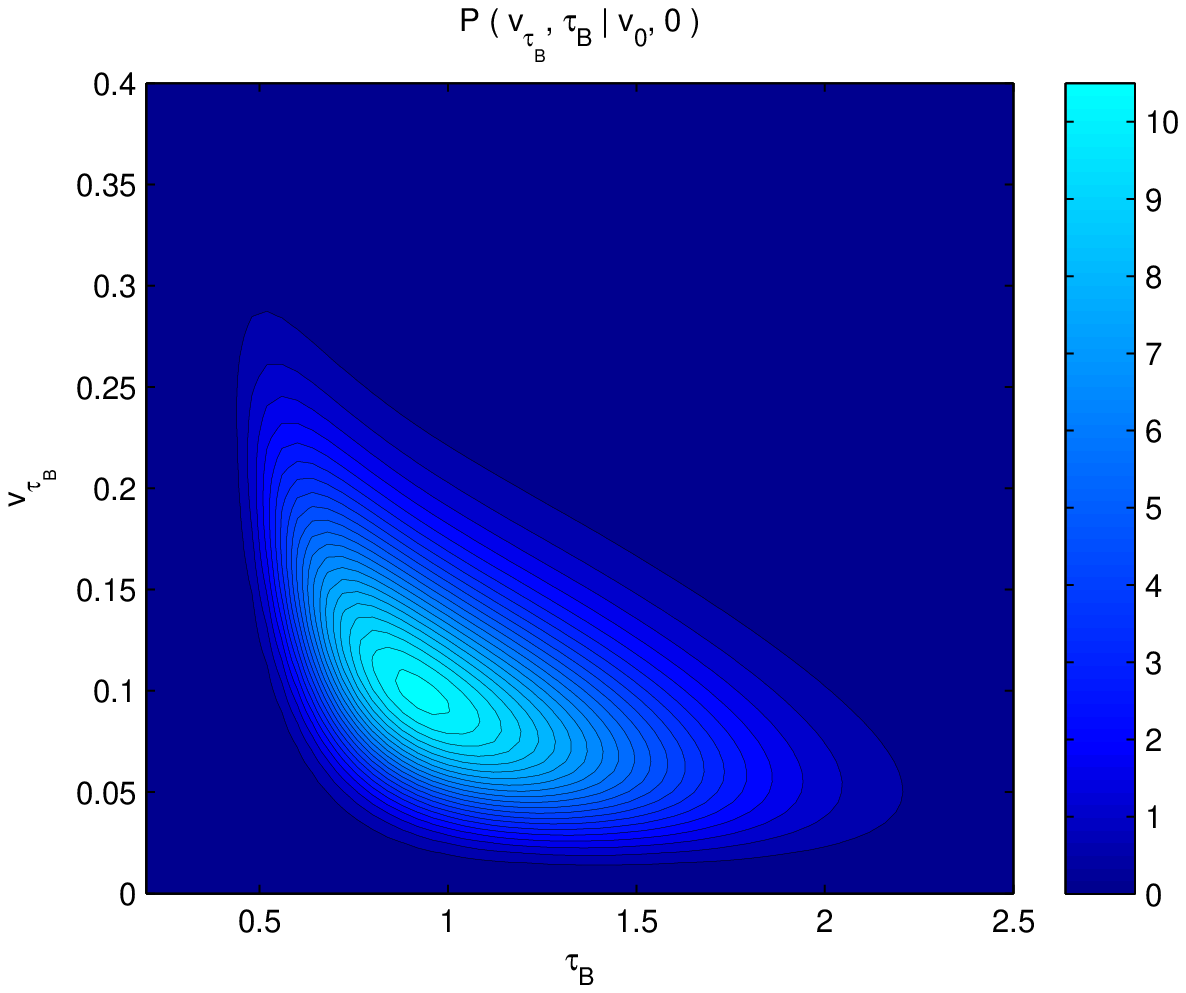}
\includegraphics[width=3in]{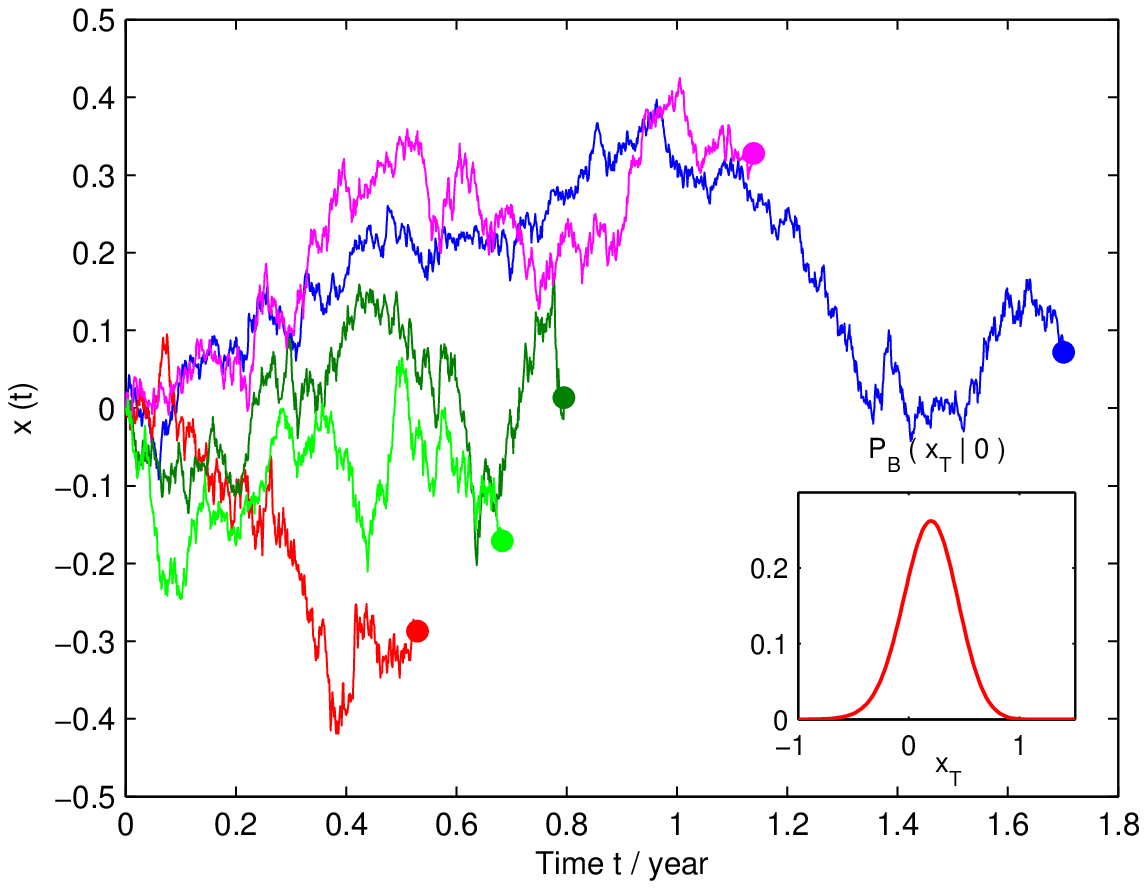}
\includegraphics[width= 3in]{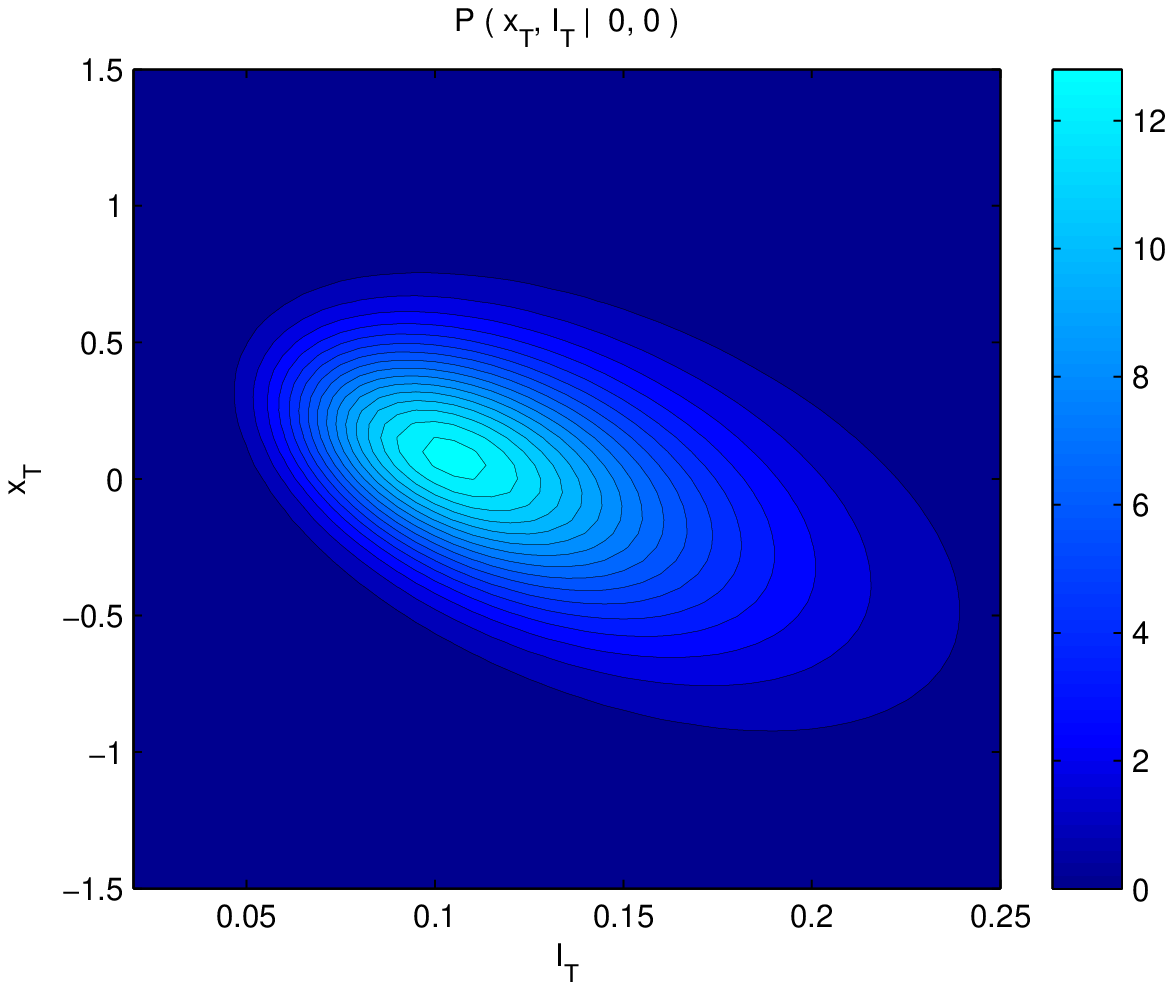}\\
\caption{(Color online, two columns) This figure is similar to figure 1, but now for the Heston stochastic volatility model. The parameters used here are: $v_0 = 0.087, \kappa = 2, \theta = 0.09, \sigma = 0.375, \mathcal{B} = v_0, r = 0.015, \rho = -0.5, T = 1.5$.}\label{fig2}
\end{figure*}

\begin{center}
\begin{table}
\scriptsize
\caption{Comparison of the analytical and the Monte Carlo (MC) simulation values for both perpetual ($\mathcal{C}_{Perp}$) and finite time-horizon ($\mathcal{C}_{Fini}$) timer call option prices under the 3/2 model. The columns indicated by RE shows the relative error (in \%). Parameters used here are: $v_0 = (0.295)^2, \kappa = 22.84, \theta = (0.4669)^2, \epsilon = 8.56, \mathcal{B} = v_0, r = 0.015, T = 1.5$.}
\label{T_1}%
\begin{tabular}{@{}|c|c|ccc|ccc|}
\hline
 \multirow{2}{*}{$K$} &  \multirow{2}{*}{$\rho$}   & \multicolumn{3}{c|}{$ \, \mathcal{C}_{Perp}$} & \multicolumn{3}{c|}{$  \,  \mathcal{C}_{Fini}$}  \\ \cline{3-8}
 & & $\,\textrm{Analytic}$ & $\textrm{MC}$ & $\textrm{RE} (\%)$  \, &   \,$\textrm{Analytic}$ & $\textrm{MC}$ & $\textrm{RE} (\%)$ \,  \\ \hline
 & - 0.5 &   \, 17.8064 & 17.8128 & -0.0359  \, & \,  17.6813 & 17.6790 & 0.0130 \,  \\
90 & 0 &  \,  17.7046 & 17.7129 & -0.0469  \, & \, 17.5385 & 17.5510 & -0.0712 \, \\
 &  0.5 & \,  17.5839 & 17.5853 & -0.0080  \, & \, 17.4260 & 17.4301 & -0.0235 \, \\  \hline
 & - 0.5 & \,  12.5780 & 12.5784 & -0.0032  \, & \, 12.4089 & 12.3998 & 0.0734 \,  \\
100 & 0 & \,  12.4619 & 12.4683 & -0.0513  \, & \, 12.2780 & 12.2890 & -0.0895 \,  \\
 & 0.5 & \,  12.3300 & 12.3231 & 0.0560  \, & \, 12.2104 & 12.2032 & 0.0590 \,  \\ \hline
 & - 0.5 & \, 8.6518 & 8.6414 & 0.0486  \, & \,  8.4381 & 8.4301 & 0.0949 \,  \\
110 & 0 & \,  8.5339 & 8.5388 & -0.0574  \, & \,  8.3531 & 8.3611 & -0.0957 \,  \\
 & 0.5 & \,  8.4026 & 8.3943 & 0.0989  \, & \,  8.3229 & 8.3153 & 0.0914 \, \\ \hline
\end{tabular}
\end{table}
\end{center}

\section{Pricing results and discussion} \label{Disc}

In the previous sections explicit formulas concerning timer options are derived for the 3/2 and the Heston model. The analytical tractability of these formulas is demonstrated by figures \ref{fig1} and \ref{fig2} and tables I and II. Figure \ref{fig1} as well as table I are devoted to the 3/2 model and figure \ref{fig2} as well as table II to the Heston model. For the 3/2 model the parameters are based on Ref. \cite{Gabriel}, where they are calibrated on market prices of S\&P500 European options to guarantee the relevance of these parameters. The parameters for the Heston model were chosen such that the two models are comparable.

The two tables compare timer option prices calculated with the formulas of the previous sections with prices obtained by Monte Carlo simulations. Results for both the perpetual and the finite time horizon timer option are presented for several strikes and correlation values. For all the Monte Carlo simulations presented here, we used 20 million samples and 3200 time steps per year. The relative error between the exact and the simulated prices is always less than $0.1\%$ confirming our formulas.

Although the prices of the timer option presented in these tables vary only slightly as a function of the correlation coefficient $\rho$, the timer option does have different features for different correlation values. This can be seen in figures \ref{fig1} and \ref{fig2}. The upper left panel shows possible realizations of the variance up to the point where the realized variance reached $\mathcal{B}$. The lower left panel shows the corresponding log-returns. In these figures we used a negative correlation. As a consequence, paths with a low log-return are more likely to have a high volatility and the corresponding option will probably be exercised sooner than an option with a high log-return. This behavior is also seen in the inset of the lower left panel. This inset shows the density given by formula (\ref{PB}) when the maximum expiry time $T$ reached $1.5$. Recall that this is the distribution of log-returns whose realized variance has not yet reached $\mathcal{B}$. Due to the negative correlation, paths with a low log-return are more likely to have reached $\mathcal{B}$ and will therefore less likely contribute to this distribution than paths with a high log-return. Therefore this distribution is clearly shifted to the right. The lower right panel shows the joint density of the log-return $x_{T}$ and the realized variance $I_{T}$ when $T$ equals $1.5$ and also illustrates this behavior.

\begin{center}
\begin{table}
\scriptsize
\caption{Comparison of the analytical and the MC simulation values for timer call option prices under the Heston model. The columns indicated by RE shows the relative error (in \%). Parameters used here are: $v_0 = 0.087, \kappa = 2, \theta = 0.09, \sigma = 0.375, \mathcal{B} = v_0, r = 0.015, \rho = -0.5, T = 1.5$.}
\label{T_1}%
\begin{tabular}{@{}|c|c|ccc|ccc|}
\hline
 \multirow{2}{*}{$K$} &  \multirow{2}{*}{$\rho$}   & \multicolumn{3}{c|}{$ \, \mathcal{C}_{Perp}$} & \multicolumn{3}{c|}{$  \,  \mathcal{C}_{Fini}$}  \\ \cline{3-8}
 & & $\,\textrm{Analytic}$ & $\textrm{MC}$ & $\textrm{RE} (\%)$  \, &   \,$\textrm{Analytic}$ & $\textrm{MC}$ & $\textrm{RE} (\%)$ \,  \\ \hline
 & - 0.5 &  \, 17.8095 & 17.7948 & 0.0826 \, & \, 17.6914 & 17.6851 & 0.0356 \,  \\
90 & 0  &  \, 17.7249 & 17.7232 & 0.0096 \, & \, 17.5351 & 17.5330 & 0.0120 \,  \\
& 0.5 & \, 17.6263 & 17.6146 & 0.0664 \, & \,  17.4627 & 17.4680 & -0.0303 \,  \\ \hline
& - 0.5 & \, 12.5789 & 12.5668 & 0.0963 \, & \, 12.4034 & 12.4010 & 0.0194 \,  \\
100 & 0 &  \, 12.4772 & 12.4763 & 0.0072 \, & \, 12.2675 & 12.2678 & -0.0024 \,  \\
& 0.5 & \,  12.3691 & 12.3586 & 0.0842 \, & \, 12.2426 & 12.2464 & -0.0310 \,  \\ \hline
& - 0.5 & \,  8.6515 & 8.6412 & 0.1192 \, & \, 8.4206 & 8.4218 & -0.0142 \,  \\
110 & 0 & \,  8.5449 & 8.5446 & 0.0035 \, & \, 8.3393 & 8.3405 & -0.0144 \,  \\
& 0.5 & \,  8.4393 & 8.4317 & 0.0890 \, & \, 8.3522 & 8.3542 & -0.0239 \,  \\ \hline
\end{tabular}
\end{table}
\end{center}

Tables I and II illustrate that the prices for timer options are quite similar for the two stochastic volatility models. Nevertheless there are important differences between the two models concerning timer options. This is illustrated by the upper panels of figures \ref{fig1} and \ref{fig2}. As already mentioned, the upper left panel shows several possible time evolutions of the variance. The inset of this panel shows the probability distribution of the stopping time $\mathcal{T}_{\mathcal{B}}$. The upper right panel shows the joint density of the variance and the stopping time, which is useful for an intuitive understanding of the time evolution of the underlying processes. The explicit form of this density is not included in the text because it is not needed to calculate prices and because it can easily be derived from expression (\ref{PXzT2}). For the 3/2 model, the probability that the variance reaches large values is larger than for the Heston model, while the probability that the variance reaches very small values is smaller than for the Heston model. Therefore the probability that the timer option will be exercised very fast is larger for the 3/2 model than for the Heston model. On the other hand for the Heston model there is a larger probability that the timer option will only be exercised after a long time than for the 3/2 model.

\section{Conclusion} \label{Conclusion}

In this paper we construct a method to price both the perpetual and the finite time-horizon timer option for a general stochastic volatility model. Pricing of such options is related to first passage time problems in that the stopping time for the option is determined by a boundary on a cumulative stochastic process. The method proposed here is based on the Duru-Kleinert time transformation and the path integral framework. Furthermore we discuss the conditions a stochastic volatility model has to satisfy in order to be able to derive closed-form pricing formulas. These general results are then applied to derive closed-form formulas for the Heston and the 3/2 stochastic volatility model. For the 3/2 model this involves the solution of the Morse potential, for the Heston model the Kratzer potential needs to be solved. Finally, our closed-form pricing formulas are shown to be computationally tractable and are validated by Monte Carlo simulation.

\appendix

\section{Path integral for Morse potential} \label{Morse}
The path integral for the Morse potential  is given in Ref. \cite{Grosche}:
\begin{align}
  & \frac{i}{\hbar} \int_0^\infty dT \, e^{i E T / \hbar} \int \limits_{ x^\prime}^{x^{\prime \prime}} \mathcal{D} x(t) \, e^{ \frac{i}{\hbar} \int_{t^\prime}^{t^{\prime \prime}} \left[ \frac{m}{2} \dot{x}^2 - \frac{\hbar^2 V_0^2}{2 m} \left( e^{2 x} - 2 \alpha \, e^x \right) \right] dt } \nonumber \\
  =& \frac{m \Gamma \left( \frac{1}{2} + \sqrt{- 2 m E} / \hbar - \alpha V_0 \right)}{\hbar^2 V_0 \Gamma \left( 1 + 2 \sqrt{- 2 m E} \right)} \, W_{\alpha V_0, \sqrt{-\frac{2m E}{\hbar^2}}} \left( 2 V_0 \, e^{x_>} \right)  \nonumber \\
  & \times  \, e^{- \left( x^\prime + x^{\prime \prime} \right) /2} \, M_{\alpha V_0, \sqrt{-\frac{2m E}{\hbar^2}}} \left( 2 V_0 \, e^{x_<} \right),
\end{align}
where $V_0 > 0$, $x_> \, (x_<)$ is the larger (smaller) of two variables $x^\prime$ and $x^{\prime \prime}$, $\Gamma(\cdot)$ is the gamma function, and $M_{\cdot, \cdot} (\cdot)$ and $W_{\cdot, \cdot} (\cdot)$ are the Whittaker functions related to the Kummer and Tricomi confluent hypergeometric functions, respectively.

This path integral (let $\hbar = 1 , - E = \Phi = \Phi_R + i \Phi_I$), analytically continued to "imaginary time" defined by
\begin{equation}
    t = - i \tau \, \, (\tau \in \mathbb{R}), \quad T = -i \Delta  \, \, (\Delta \in \mathbb{R}),
\end{equation}
results in:
\begin{eqnarray}
%  & & i \int_0^{- i \infty} d(- i \Delta) \, e^{i (-\Phi) (- i \Delta)}  \int \limits_{x(0) = x^\prime}^{x(- i \Delta) = x^{\prime \prime}} \mathcal{D} x(- i \tau) \nonumber \\
%  && \quad \quad \times e^{ i \int_{0}^{- i \Delta} \left[ \frac{m}{2} \left( \frac{dx}{d\tau} \frac{d\tau}{dt} \right)^2 - \frac{V_0^2}{2 m} \left( e^{2 x} - 2 \alpha \, e^x \right) \right] d(-i \tau) } \nonumber  \\
  && \int_0^\infty d\Delta \, e^{- \Phi \Delta} \int \limits_{x^\prime}^{x^{\prime \prime}} \mathcal{D} x(\tau)  \, e^{- \int_{0}^{\Delta} \left[ \frac{m}{2} \left( \frac{dx}{d\tau}  \right)^2 + \frac{V_0^2}{2 m} \left( e^{2 x} - 2 \alpha \, e^x \right) \right] d \tau } \nonumber  \\
  &=& \frac{m \Gamma \left( \frac{1}{2} + \sqrt{2 m \Phi} - \alpha V_0 \right)}{V_0 \Gamma \left( 1 + 2 \sqrt{2 m \Phi} \right)} \, e^{- \left( x^\prime + x^{\prime \prime} \right) /2} \nonumber \\
  && \times W_{\alpha V_0, \sqrt{2m \Phi}} \left( 2 V_0 \, e^{x_>} \right) \, M_{\alpha V_0, \sqrt{2m \Phi}} \left( 2 V_0 \, e^{x_<} \right).
\end{eqnarray}
Therefore the inverse Laplace transform gives the expression of the propagator evolving in the "imaginary" time horizon $\left[ 0, \Delta \right]$ as shown in Ref. \cite{Kleinert}:
\begin{eqnarray}
  &&  \int \mathcal{D} x(\tau)  e^{- \int_{0}^{\Delta} \left[ \frac{m}{2} \left( \frac{dx}{d\tau}  \right)^2 + \frac{V_0^2}{2 m} \left( e^{2 x} - 2 \alpha \, e^x \right) \right] d \tau } \nonumber  \\
  &=&  \int_{\Phi_R - i \infty}^{\Phi_R + i \infty} \frac{d \Phi}{2 \pi i}  \, e^{\Phi \Delta} \frac{m \Gamma \left( \frac{1}{2} + \sqrt{2 m \Phi} - \alpha V_0 \right)}{V_0 \Gamma \left( 1 + 2 \sqrt{2 m \Phi} \right)} \, e^{- \frac{ x^\prime + x^{\prime \prime}}{2}} \nonumber \\
  && \times W_{\alpha V_0, \sqrt{2m \Phi}} \left( 2 V_0 \, e^{x_>} \right) \, M_{\alpha V_0, \sqrt{2m \Phi}} \left( 2 V_0 \, e^{x_<} \right) \nonumber \\
  &=& 2m \int_0^{\infty} \frac{d \Phi_I}{\pi} \, \textrm{Re} \left[ \, e^{\Phi \Delta}  \int_0^\infty \frac{d \xi}{\sinh \xi} \, e^{2 \alpha V_0 \xi} \right. \nonumber \\
  && \times \left. \, e^{- V_0 \left( e^{x^\prime} + e^{x^{\prime \prime} }\right) \coth \xi} \, I_{2 \sqrt{2 m \Phi}} \left( \frac{2 V_0 \, e^{\frac{x^\prime + x^{\prime \prime}}{2}}}{\sinh \xi} \right) \right],
\end{eqnarray}
under the condition
\begin{equation}
    \textrm{Re} \left[ \frac{1}{2} + \sqrt{2 m \Phi} - \alpha V_0 \right] > 0 ,
\end{equation}
that is
\begin{equation}\label{Restriction2}
    \Phi_R > \frac{\left( \textrm{Re} [\alpha] V_0 - \frac{1}{2} \right)^2}{2m}.
\end{equation}

\section{Path integral for Kratzer potential}\label{Kratzer}
The path integral for Kratzer potential is given in Ref. \cite{Grosche} $\left( \lambda > 0, \kappa = \frac{e_1 e_2}{\hbar} \sqrt{- \frac{m}{2E}} \right)$:
\begin{align}
   & \frac{i}{\hbar} \int_0^\infty dT \, e^{i E T / \hbar} \int\limits_{ r^\prime}^{ r^{\prime \prime}} \mathcal{D} r(t) \, e^{ \frac{i}{\hbar} \int_{t^{\prime}}^{t^{\prime \prime}}  \left( \frac{m}{2} \dot{r}^2 + \frac{e_1 e_2}{r} - \frac{\hbar^2}{2m} \frac{\lambda^2 - \frac{1}{4}}{r^2} \right) dt } \nonumber \\
   =& \frac{1}{\hbar} \sqrt{- \frac{m}{2E}} \frac{\Gamma \left( \frac{1}{2} + \lambda - \kappa \right)}{\Gamma \left( 2 \lambda + 1 \right)} \nonumber  \\
   & \times W_{\kappa, \lambda} \left( \sqrt{- 8 m E} \, \frac{r_>}{\hbar} \right) \, M_{\kappa, \lambda} \left(\sqrt{- 8 m E} \, \frac{r_<}{\hbar} \right),
\end{align}

The analytical continuation of this path integral (let $\hbar = 1 ,m = 1, e_1 e_2 = \beta, - E =  \Phi = \Phi_R + i \Phi_I$) as used in Appendix \ref{Morse} gives $\left( \kappa = \frac{\beta}{\sqrt{2 \Phi}} \right)$:
\begin{align}
%  && i \int_0^{- i \infty} d (- i \Delta) \, e^{i (- \Phi) (- i \Delta) } \int\limits_{r(0) = r^\prime}^{r(- i \Delta) = r^{\prime \prime}} \mathcal{D} r(- i \tau) \nonumber \\
%  && \quad \quad \times \, e^{ i \int_0^{- i \Delta}  \left( \frac{m}{2} \left( \frac{d r}{d \tau} \frac{d \tau}{d t} \right)^2 + \frac{\beta}{r} - \frac{1}{2m} \frac{\lambda^2 - \frac{1}{4}}{r^2} \right) d (- i \tau) } \nonumber \\
  & \int_0^\infty d \Delta \, e^{- \Phi \Delta} \int\limits_{r(0) = r^\prime}^{r(\Delta) = r^{\prime \prime}} \mathcal{D} r(\tau) e^{ - \int_0^{\Delta}  \left[ \frac{1}{2} \left( \frac{dr}{d\tau}  \right)^2  +  \frac{\lambda^2 - \frac{1}{4}}{2 r^2} - \frac{\beta}{r} \right] d\tau } \nonumber \\
  =&  \frac{\Gamma \left( \frac{1}{2} + \lambda - \kappa \right)}{\sqrt{2 \Phi} \, \Gamma \left( 2 \lambda + 1 \right)} W_{\kappa, \lambda} \left( 2 \sqrt{ 2 \Phi} \, \, r_> \right) \, M_{\kappa, \lambda} \left(2 \sqrt{2 \Phi} \, \, r_< \right).
\end{align}
Therefore the inverse Laplace transform leads to the propagator in time horizon $[0, \Delta]$ as in Ref. \cite{Grosche2}:
\begin{eqnarray}
  && \int \mathcal{D} r(\tau) e^{ - \int_0^{\Delta}  \left[ \frac{1}{2} \left( \frac{dr}{d\tau}  \right)^2  +  \frac{\lambda^2 - \frac{1}{4}}{2 r^2} - \frac{\beta}{r} \right] d\tau } \nonumber \\
  &=&  \int_0^\infty \frac{d \Phi_I}{\pi} \, \textrm{Re} \left[ \, e^{\Phi \Delta} \, 2 \sqrt{2 \Phi} \int_0^\infty  d \xi  \right. \nonumber \\
  && \times  \, e^{2 \beta \xi - \sqrt{2 \Phi} \left( r_> + r_< \right) \coth \left( \sqrt{2 \Phi} \, \xi \right) } \nonumber \\
  && \times \left. \frac{\sqrt{r_> \, r_<}}{\sinh \left( \sqrt{2 \Phi} \, \xi \right)} \, \, I_{2 \lambda} \left( \frac{2 \sqrt{2 \Phi} \, \sqrt{r_> \, r_<} }{\sinh \left( \sqrt{2 \Phi} \, \xi \right)} \right) \right],
\end{eqnarray}
valid under the condition
\begin{equation}
    \textrm{Re} \left[ \frac{1}{2} + \lambda - \frac{\beta}{\sqrt{2 \Phi}} \right] > 0,
\end{equation}
that is
\begin{equation} \label{Restriction1}
    \Phi_R >  \frac{1}{2} \left( \frac{\beta}{\lambda + \frac{1}{2}} \right)^2.
\end{equation}

\bibliography{bibfile}

\end{document}